\documentclass[twoside]{article}
\usepackage{amsfonts,amssymb,amsbsy,textcomp,marvosym,picins,amsmath,caption,threeparttable,amsthm,subfigure}
\usepackage{eurosym,mathrsfs,fancyhdr,CJK,multicol,graphics,indentfirst,color,bm,upgreek,booktabs,graphicx}

\renewcommand{\thefootnote}{\fnsymbol{footnote}}
\def\footnoterule{\kern 1mm \hrule width 10cm \kern 2mm}

\allowdisplaybreaks
\catcode`@=11
\def\title#1{\vspace{3mm}\begin{flushleft}\vglue-.1cm\Large\bf\boldmath\protect\baselineskip=18pt plus.2pt minus.1pt #1
\end{flushleft}\vspace{1mm} }
\def\author#1{\begin{flushleft}\normalsize #1\end{flushleft}\vspace*{-4pt} \vspace{3mm}}
\def\address#1#2{\begin{flushleft}\vglue-.35cm${}^{#1}$\small\it #2\vglue-.35cm\end{flushleft}\vspace{-2mm}\par}

\catcode`@=11
\def\section{\@startsection{section}{1}{\z@}%
 {-3ex \@plus -.3ex \@minus -.2ex}%
 {2.2ex \@plus.2ex}%
{\normalfont\normalsize\protect\baselineskip=14.5pt plus.2pt minus.2pt\bfseries}}
\def\subsection{\@startsection{subsection}{2}{\z@}%
 {-3ex\@plus -.2ex \@minus -.2ex}%
 {2ex \@plus.2ex}%
{\normalfont\normalsize\protect\baselineskip=12.5pt plus.2pt minus.2pt\bfseries}}
\def\subsubsection{\@startsection{subsubsection}{3}{\z@}%
 {-2.2ex\@plus -.21ex \@minus -.2ex}%
 {1.4ex \@plus.2ex}
{\normalfont\normalsize\protect\baselineskip=12pt plus.2pt minus.2pt\sl}}

\usepackage{graphicx}
\usepackage{amsmath}
\usepackage{bm}
\usepackage[subnum]{cases}
\usepackage{multirow}
\usepackage{booktabs}
\usepackage{color}
\usepackage{subfig}
\usepackage{url}
\usepackage{enumitem}
\usepackage[export]{adjustbox}
\usepackage{cuted}
\newcommand{\tabincell}[2]{\begin{tabular}{@{}#1@{}}#2\end{tabular}}
\usepackage[super,square,sort&compress]{natbib}
\usepackage{tikz}

\renewcommand\thefootnote{\protect\circled{\arabic{footnote}}}

\begin{document}
\begin{CJK*}{GBK}{song}
\thispagestyle{empty}
\vspace*{-13mm}
\noindent {\small Wang X, Huang C, Yao L {\it et al.} A survey on expert recommendation in community question answering.
JOURNAL OF COMPUTER SCIENCE AND TECHNOLOGY \ 33(1): \thepage--\pageref{last-page}
\ January 2018. DOI 10.1007/s11390-015-0000-0}
\vspace*{2mm}

\title{A Survey on Expert Recommendation in Community Question Answering}

\author{
Xianzhi Wang$^1$, \textit{Member}, \textit{ACM}, \textit{IEEE}, Chaoran Huang$^2$, Lina Yao$^2$, \textit{Member}, \textit{ACM}, \textit{IEEE},\\ Boualem Benatallah$^2$, \textit{Member}, \textit{IEEE},
and Manqing Dong$^2$, \textit{Student Member}, \textit{ACM}, \textit{IEEE}
}

\address{1}{
School of Software, University of Technology Sydney, NSW 2007, Australia
}
\address{2}{
School of Computer Science and Engineering, University of New South Wales, Sydney, 2052 NSW, Australia
}

\vspace{2mm}

\noindent E-mail: xzwang@smu.edu.sg, \{chaoran.huang, lina.yao, b.benatallah, manqing.dong\}@unsw.edu.au\\[-1mm]

\noindent Received July 15, 2018; revised October 14, 2018.\\[1mm]

\let\thefootnote\relax\footnotetext{{}\\[-4mm]\indent\quad Survey
\\[.5mm]\indent\quad \copyright 2018 Springer Science\,+\,Business Media, LLC \& Science Press, China}

\noindent {\small\bf Abstract} \quad  {\small Community question answering (CQA) represents the type of Web applications where people can exchange knowledge via asking and answering questions. One significant challenge of most real-world CQA systems is the lack of effective matching between questions and the potential good answerers, which adversely affects the efficient knowledge acquisition and circulation. On the one hand, a requester might experience many low-quality answers without receiving a quality response in a brief time; on the other hand, an answerer might face numerous new questions without being able to identify their questions of interest quickly. Under this situation, expert recommendation emerges as a promising technique to address the above issues. Instead of passively waiting for users to browse and find their questions of interest, an expert recommendation method raises the attention of users to the appropriate questions actively and promptly. The past few years have witnessed considerable efforts that address the expert recommendation problem from different perspectives. These methods all have their issues that need to be resolved before the advantages of expert recommendation can be fully embraced. In this survey, we first present an overview of the research efforts and state-of-the-art techniques for the expert recommendation in CQA. We next summarize and compare the existing methods concerning their advantages and shortcomings, followed by discussing the open issues and future research directions.}

\vspace*{3mm}


\vspace*{4mm}

\end{CJK*}
\baselineskip=13.2pt plus.2pt minus.2pt
\parskip=0pt plus.2pt minus0.2pt
\begin{multicols}{2}

\section{Introduction}
\label{intro}

The prosperity of crowdsourcing and web 2.0 has fostered numerous online communities featuring question answering (Q\&A) activities.
Such communities exist in various forms such as dedicated websites, online forums, and discussion boards.
They provide a venue for people to share and obtain knowledge by asking and answering questions, known as community question answering (CQA)~\cite{srba2016comprehensive}.
While traditional online information seeking approaches (e.g., search engines) retrieve information from existing information repositories based on keywords, they face several challenges.
First, answers to some questions may not exist in the previously answered questions~\cite{liu2012web} and thus cannot be retrieved from existing repositories directly.
Second, most real-world questions are written in complicated natural languages that require certain human intelligence to be understood.
Third, some questions inherently seek people's opinions and can only be answered by humans. 
While machines find difficult to handle the above cases, CQA can leverage the ``wisdom of crowds" and obtain answers from multiple people simultaneously.
Typical Q\&A websites include Yahoo! Answers (answers.yahoo.com),
Quora (www.quora.com),
and Stack Overflow (stackoverflow.com).
The first two websites cover a wide range of topics, while the last only focuses on the topic of computer programming.

Though advantages over the traditional information seeking approaches, CQA faces several unique challenges.
First, a CQA website may have tens of thousands of questions posed every day, let alone the millions of questions that already exist on the website. The huge volume of questions makes it difficult for a general answerer to find the appropriate questions to answer~\cite{guo2008tapping}.
Second, answerers usually have varying interest and expertise in different topics and knowledge domains. Thus, they may give answers of varying quality to different questions. The time required for preparing answers~\cite{su2007internet} and the intention of answering also affect the quality of their responses. 
An extreme case is that answerers may give irrelevant answers that distract other users~\cite{agichtein2008finding} without serious thinking.
All the above situations cause additional efforts of an information seeker in obtaining good answers.
Third, instead of receiving an answer instantly, users in CQA may need to wait a long time until a satisfactory answer appears.
Previous studies~\cite{li2010routing} show that many questions on real-world CQA websites cannot be resolved adequately, meaning the requesters recognize no best answers to their questions within 24 hours.

Fortunately, several studies~\cite{fisher2006you,viegas2004newsgroup,welser2007visualizing} have shown that some core answerers are the primary drivers of answer production in the many communities. Recent work on Stack Overflow and Quora~\cite{anderson2012discovering} further indicates that these sites consist of a set of highly dedicated domain experts who aim at satisfying requesters' query but more importantly at providing answers with high lasting value to a broader audience.
All these studies suggest the needs for recommending a small group of most competent answerers, or experts to answer the new questions.
In fact, the long-tail phenomena in many real-world communities, from the statistic perspective, lays the ground of the rationale of expert recommendation in CQA~\cite{adamic2008knowledge}, as most answers and knowledge in the communities come from only a minority of users~\cite{movshovitz2013analysis,adamic2008knowledge}.
As an effective means of addressing the practical challenges of traditional information seeking approaches, expert recommendation methods bring up the attention of only a small number of experts, i.e., the users who are most likely to provide high-quality answers, to answer a given question~\cite{yimam2003expert}.
Since expert recommendation inherently encourages fast acquisition of higher-quality answers, it potentially increases the participation rates of users, improves the visibility of experts, as well as fosters stronger communities in CQA.

Given the advantages of expert recommendation and related topics such as question routing~\cite{li2010routing,zhou2009routing} and question recommendation~\cite{qu2009probabilistic} in the domains of Natural Language Processing (NLP) and Information Retrieval (IR), we aim to present a comprehensive survey on the expert recommendation in CQA.
On the one hand, considerable efforts have been conducted on the expert recommendation and have delivered fruitful results.
Therefore, it is necessary to review the related methods and techniques to gain a timely and better understanding of state of the art.
On the other hand, despite the active research in CQA, expert recommendation remains a challenging task. For example, the sparsity of historical question and answer records, low participation rates of users, lack of personalization in recommendation results, the migration of users in or out of communities, and lack of comprehensive consideration of different clues in modeling users expertise are all regarded as challenging issues in literature.
Given the diverse existing methods, it is crucial to develop a general framework to evaluate these methods and analyze their shortcomings, as well as to point out promising future research directions.

To the best of our knowledge, this is the first comprehensive survey that focuses on the expert recommendation issue in CQA.
The remainder of the article is organized as follows.
We overview the expert recommendation problem in Section~\ref{Problem} and its current applications in CQA in Section~\ref{application}.
In Section~\ref{Methods}, we present the classification and introduction of state of the art expert recommendation methods.
In Section~\ref{comparison}, we compare the investigated expert recommendation methods on various aspects and discuss their advantages and pitfalls.
In Section~\ref{Directions}, we highlight several promising research directions.
Finally, we offer some concluding remarks in Section~\ref{conclusion}.

\section{Expert Recommendation Problem}
\label{Problem}

\begin{figure*}[!htb]
\centering
\includegraphics[width=9.45cm]{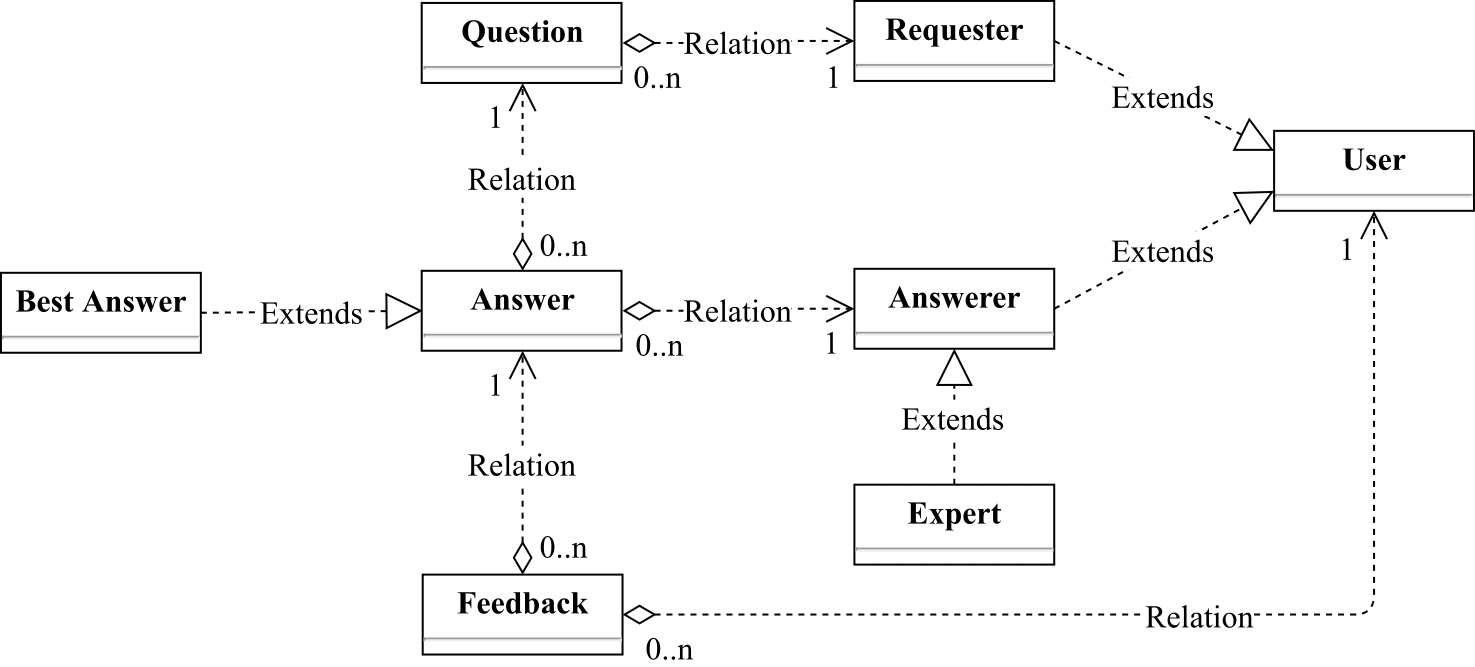}
\caption{Elements of expert recommendation in CQA.}
\label{Fig1}
\end{figure*}

The expert recommendation issue is also known as the question routing or expert finding problem.
The basic inputs of an expert recommendation problem include users (i.e., requesters and answerers) and user-generated content (i.e., the questions raised by requesters and the answers provided by answerers).
More inputs might be available depending on the application scenarios. Typically, they include user profiles (e.g., badges, reputation scores, and links to external resources such as Web pages), users' feedback on questions and answers (e.g., textual comments and votings), and question details (e.g., the categories of questions and duplication relations among questions).
The relationship among the different types of inputs of an expert recommendation problem is described in the class diagram shown in Fig.~\ref{Fig1}.


\baselineskip=13.2pt plus.2pt minus.2pt
\parskip=0pt plus.2pt minus0.2pt


Question answering websites usually organize information in the form of threads. Each thread is led by a single question, which is replied to with none, one, or multiple answers.
Each question or answer is provided by a single user, called a requester or an answerer, respectively.
A requester may ask multiple questions, and each answerer may answer various questions.
A user can be either a requester or an answerer, or both at the same time in the same CQA website, and all users are free to provide different types of feedback on the posted questions and answers.
For example, in \textit{Stack Overflow}, any registered user can comment and vote (by giving a thumb up or thumb down) on an answer posted for any question, and the requester has the authority to mark one from the posted answers as the best answer. In case that the requester has not designated the best answer within a specified period, the system will automatically mark the response that received the highest voting score as the best answer.

The objective of the expert recommendation problem is to raise the attention of experts, i.e., a small number of users who are most likely to provide high-quality answers, to the given question based on the above problem inputs.
Despite the various possible types of inputs, only a subset of them might be available in a specific application scenario. Therefore, researchers may define the expert recommendation problem differently according to the inputs.
Besides, researchers may take into account different concerns and expect different types of outputs from their methods.
Generally, topical relevance and expertise are the two most considered aspects of concerns by the existing research. While some researchers develop methods to find a group of high-quality answerers, other researchers aim to deliver a ranked list, where the users are ranked according to their potential to provide the best answer.
We will elaborate the variations in the problem definition in Section~\ref{comparison}.

Generally, it is only necessary to recommend experts when the new question is significantly different from any previous questions with best answers, meaning that no satisfactory answers are readily available within the archive of best answers to the earlier questions.
Expert recommendation generally brings about the following advantages to CQA:
i) users usually prefer answers from experts, who are supposed to have sufficient motivation and knowledge to answer the given questions and therefore more likely to provide high-quality answers promptly;
ii) expert recommendations can potentially reduce the waiting time of requesters in finding satisfactory answers as well as the time of experts in finding their questions of interests;
iii) by bridging the gap between requesters and answerers, expert recommendations can potentially promote their participation rates and thus foster stronger communities. Since experts are recommended with questions that fit their expertise, their visibility is expected to be improved as well.

\section{Current Applications in CQA}
\label{application}

Currently, there exist various Q\&A websites where expert recommendation techniques are applied or can be potentially applied.
Due to the large number of Q\&A websites that exist nowadays, we selectively list some typical Q\&A websites by launch year in Table~\ref{Table1}.
In the following subsections, we will categorize and give further illustrations of several typical websites of each category.


\tabcolsep 9pt
\renewcommand\arraystretch{1.3}
\begin{table*}[!htb]
\centering
\caption{\label{Table1} Some Popular Question Answering Communities}\vspace{-2mm}
{\footnotesize
\begin{tabular*}{\linewidth}{cccccc}
\hline\hline\hline
Community & Language & Specialized Domain & Launch Year & Still Active & Quality Guarantee\\
\hline
MedHelp                 & English& Medical     & 1994    & Y  & Y       \\
Mad Scientist Netwok    & English& Various     & 1995    & Y  & Y       \\
WebMD                   & English& Medical     & 1996    & Y  & Y       \\
Google Answers          & Multiple& Various    & 2002    & N  & Y       \\
Naver KiN               & Korean & Various     & 2002    & Y  & N       \\
WikiAnswers             & English& Various     & 2002    & Y  & N       \\
Answerbag               & English& Various     & 2003    & Y  & N       \\
IAsk                    & Chinses& Various     & 2005    & Y  & N       \\
Baidu Knows             & Chinese& Various     & 2005    & Y  & N       \\
Live QnA                & English& Various     & 2006    & N  & N       \\
TurboTax Live Community & English& Tax           & 2007    & Y  & N       \\
Sogou Wenwen            & Chinese& Various     & 2007    & Y  & N       \\
Stack Overflow          & English& Programming & 2008    & Y  & N       \\
Quora                   & English& Various     & 2010    & Y  & N       \\
Seasoned Advice         & English& Cooking     & 2010    & Y  & N       \\
\hline\hline\hline
\end{tabular*}
}
\end{table*}
\baselineskip=13.2pt plus.2pt minus.2pt
\parskip=0pt plus.2pt minus0.2pt

\subsection{Early CQA Services}

Most early-stage Q\&A services (e.g., the first four websites in Table~\ref{Table1}) meet a requesters' information needs by resorting to the opinions of experts rather than the crowd. These experts are acknowledged by either the websites or third-party authorities and are often limited in number. They usually have rich knowledge and experience in some domains but require a payment for the answers they provide. We introduce two of these websites as examples as follows:

\textit{Mad Scientist Network}\footnote{http://www.madsci.org/, May 2018.}: a famous ask-a-scientist web service where people ask questions by filling forms and moderators are responsible for reviewing the questions and sending them to the appropriate members for answers. The moderators will also review the answers before making them public.

\textit{Google Answers}\footnote{http://answers.google.com/, May 2018.}: a knowledge market service designed as an extension to Google's search service. There were a group of answerers called Google Answers Researchers who are officially approved to answer questions through an application process.
Instead of passively waiting for other people to moderate or answer their questions, people can actively find the potential answerers by themselves and pay the answerers. 

\subsection{General-purpose CQA Websites}

The Q\&A services that emerge in the past two decades are increasingly leveraging the ``wisdom of the crowd'' rather than a small number of experts to give answers.
Websites following this philosophy allow any users to voluntarily answer any questions on their free will and most of them serve as general purpose platforms for knowledge sharing rather then domain focused ones.
We overview some typical general purpose websites as follows:

\textit{Quora}: one of the largest existing Q\&A website where users can ask and answer questions, rate and edit the answers posted by others.

\textit{Zhihu}\footnote{http://www.zhihu.com/, May 2018.}: a Chinese Q\&A website similar to Quora. It allows users to create and edit questions and answers, rate system, and tag questions. Also, users may also post blogs in Zhihu for sharing while others can view and comment on such posts.

\textit{Naver KiN}\footnote{http://kin.naver.com/, May 2018.}: a Korean CQA community, one of the earlier cases of expansion of search service using user-generated content. 

\textit{WikiAnswers}\footnote{http://www.wikianswers.com/, May 2018.}: a wiki service that allows people to raise and answer questions, as well as edit existing answers to questions. It uses a so-called ``alternates system" to automatically merge similar questions. Since an answer may be associated with multiple questions, duplicated entries can be avoided to some extent.

\textit{Answerbag}\footnote{http://www.answerbag.com/, May 2018.}: a CQA community where users can ask and answer questions, give comments to answers, rate questions, rate answers, and suggest new categories.

\textit{Live QnA}\footnote{http://qna.live.com/, May 2018.}: also known as MSN QnA, was part of Microsoft MSN group services. In this system, users can ask and answer questions, tag them to specific topics, and gain points and reputations by answering questions.

\subsection{Domain-focused CQA Websites}

Compared with those general purpose websites, each domain-focused Q\&A website only covers limited topics or knowledge domains.
The Stack Exchange networks are probably the largest host of domain-focused Q\&A websites nowadays.
Some typical websites hosted by it include the following:

\textit{MathOverflow}\footnote{http://mathoverflow.net/, May 2018.}: a Q\&A website focused on mathematical problems.

\textit{AskUbuntu}\footnote{http://askubuntu.com/, May 2018.}: a website supporting Q\&A activities related to Ubuntu operation Systems.

\textit{StackOverflow}: a Q\&A website focused on computer programming.

All these websites follow similar sets of styles and functions.
Apart from the basic question answering features, they commonly use badges to recognize the achievement of answerers and grant badges to users based on their reputation points.
Users can also unlock more privileges with higher reputation points.

\subsection{Summary}
In summary, despite the prevalence of diverse types of Q\&A websites, few of them have incorporated any effective expert recommendation techniques to bridge requesters and answers.
To the best of our knowledge, currently, the only implementation of the idea of routing questions to the appropriate users in Q\&A is called ``Aardvark"~\cite{horowitz2010anatomy}.
However, the primary purpose of this system is to serve as an enhanced search engine, and the expert recommendation techniques it employs are still at a preliminary stage.
Recently, Bayati et al.~\cite{bayati2016security} design a framework for recommending security experts for software engineering projects.
This framework offers more strength to facilitate expert recommendation by considering multiple aspects of users such as programming language, location, and social profiles on dominant programming Q\&A websites like StackOverflow.
Since the Q\&A systems can be regarded as a type of crowdsourcing systems~\cite{rjab2016characterization}, the expert recommendation methods for a Q\&A system can potentially be generalized and applied to general crowdsourcing systems as well. 

\section{Expert Recommendation Methods}
\label{Methods}

As the major technique to facilitate effective CQA, considerable efforts have been contributed to the expert recommendation research from the information retrieval (IR), machine learning, and social computing perspectives, and have delivered fruitful results.
We classify the state of the art expert recommendation methods into eight categories and review the methods by category in the following subsections.

\subsection{Simple Methods}

One of the most critical tasks of expert recommendation is to evaluate users.
Given a new question to be answered, some methods use simple metrics such as counts of positive/negative votes, proportions of best answers, and the similarity between the new question and users' previous answered questions to evaluate users' fitness to answer the questions.
In the following, we introduce the methods that use the three metrics, respectively. For any of these methods, a higher score indicates a better answerer.

\textit{Votes}: the method evaluates a user by the number of affirmative votes minus the number of negative votes, combined with the total percentage of affirmative votes that the user receives from other users averaged over all the answers the user have attempted.

\textit{Best answer proportion}: this method ranks users by the fraction of best answers among all the answers attempted by an answerer. The best answers are either awarded by the requester of questions or by the question answering platform when requesters designate no best answers.

\textit{Textual similarity}: the most famous method for measuring textual similarity is to compute the cosine similarity based on the term frequency-inverse document frequency (TF-IDF) model, a classic vector space model (VSM)~\cite{baeza1999modern} borrowed from the information retrieval domain.
VSM is readily applicable to computing the similarity of an answerer's profile to a given question. Therefore, it can be directly used for the expert recommendation by relating a new question to the answerers who have previously answered the most relevant questions to the given question.

\subsection{Language Models}

Despite the simplicity, VSM adopts the ``bag-of-words" assumption and thus brings the high-dimension document representation issue.
In contrast, language models use a generative approach to compute the word-based relevance of a user's previous activities to the given question, and in turn, to predict the possibility of a user answering the question.
Such models can, to some extent alleviate the high dimension issue.
In a language model, the users whose profiles are most likely to generate the given question are believed to have to highest probability to answer the given question.
The model finally returns a ranked list of users according to their likelihood of answering the given question.

The language model-based methods include profile-based methods and document-based methods. The former~\cite{balog2006formal} models the knowledge of each user with the associated documents and ranks the candidate experts for a given topic based on the relevance scores between their profiles and the given question. The latter~\cite{balog2006formal} finds related documents for a given topic and ranks candidates based on mentions of the candidates in the related documents.

\subsubsection{QLL and Basic Variants}

Among the methods of this category, query likelihood language (QLL) model~\cite{miller1999hidden} is the most popular technique. QLL calculates a probability that user profiles will generate terms of the routed question.
The traditional language models often suffer the mismatch between the question and user profiles caused by the co-occurrence of random words in user profiles or questions resulting from data sparseness. Translation models~\cite{zhou2012joint} overcomes data sparseness by employing statistical machine translation and can differentiate between exact matched words and translated semantically related ones.
A typical work~\cite{liu2005finding} using this method views the problem as an IR problem. It considers the new question as a query and the expert profiles as documents. It next estimates an answerer's expertise by combining its previously answered questions, and regards experts as the users who have answered the most similar questions in the past.

Besides the basic models, many variants of QLL have also emerged as alternatives or enhancements.
For example, 
Liu et al. propose two variants of the basic language model, namely \textit{relevance-based language model}~\cite{lavrenko2001relevance} and \textit{cluster-based language model}~\cite{liu2004cluster} to rank user profiles.
Petkova and Croft~\cite{petkova2008hierarchical} propose a hierarchical language model which uses a finer-grained approach with a linear combination of the language models built on subcollections of documents.

\subsubsection{Category-sensitive QLL}

Considering the availability of categories in many Q\&A websites, Li et al.~\cite{li2011question} propose a category-sensitive QLL model to exploit the hierarchical category information presented with questions in Yahoo! Answers. Once a question gets categorized, the task is to find the users who are most likely to answer that question within its category. Their experiments over the Yahoo! Answers dataset show that taking categories into account improves the recommendation performance.
A limitation of the category-sensitive model is that categories need to be well predefined and some questions might be closely related to multiple categories due to the existence of similar categories that share the same contexts.
A possible solution to address this limitation is the transferred category-sensitive QLL model~\cite{li2011question}, which additionally builds and considers the relevance between categories.

\subsubsection{Expertise-aware QLL}

Zheng et al.~\cite{zheng2012algorithm} linearly combine two aspects, user relevance (computed based on the QLL) and answer quality (estimated using a maximum entropy model), using the simple weighted sum method to represent user expertise on a given question.
Besides the relevance and quality aspects, Li et al.~\cite{li2010routing} further consider the availability of users and use the weighted sum of the three aspects to represent user expertise on a given question.
In particular, the relevance is estimated using the QLL model,
the answer quality is estimated as the weighted average of previous answer quality incorporated with the Jelinek-Mercer smoothing~\cite{zhai2004study} method, and users' availability to answer a given question during a given period is predicted by an autoregressive model.
Compared with most existing methods, this method exploits not only time series availability information of users but also multiple metadata features such as answer length, question-answer length, number of answers for this question, the answerer's total points, and the answerer's best answer ratio.
These features have rarely been used by the existing research.

\subsection{Topic Models}

Since language models are based on exact word matching, they are most effective when they are used within the same topic. Besides, they are not able to capture more advanced semantics and solve the problem of the lexical gap between a question and user profiles.
In contrast, topic models do not require the word to appear in the user profile, as it measures their relationship in the topic space rather than in the word space.
It can, therefore, alleviate the lexical gap problem and previous experimental evaluations have confirmed the better performance of many topic models over language models~\cite{liu2010predicting,riahi2012finding}.
Here, we focus on reviewing two most widely used topic models, Probabilistic Latent Semantic Analysis (PLSA) and \textit{Latent Dirichlet Allocation (LDA)}, as well as their variants and a few other models.

\subsubsection{PLSA and Its Variants}

Probabilistic Latent Semantic Analysis (PLSA) a.k.a. Probabilistic Latent Semantic Indexing (PLSI)~\cite{hofmann1999probabilistic} is developed based on Latent Semantic Indexing (LSI)~\cite{deerwester1990indexing}, which uses Singular Value Decomposition to represent a document in a low-dimension space.
Compared with LSI, which lacks semantic explanation, PLSA uses latent topics to represent documents and model the data generation process as a Bayesian network.
In this way, it can leverage the semantic between words in documents to reduce the document representation space dimension.
There are generally two classes of PLSA-based methods that model users directly and indirectly, respectively. We briefly review the two classes of methods as follows:

\textit{Direct User Model by PLSA}. Methods of this class treat all the questions that a user accesses as one document. Then, PLSA is used directly to derive the topic information of the user using word distributions.
A typical method of this class~\cite{qu2009probabilistic} would identify the underlying topics of questions to match users' interest and thereby help the capable users locate the right questions to answer.
The Expectation Maximization (EM) algorithm is generally used to find a local maximum of the log-likelihood of the question collection and to learn model parameters.

\textit{Indirect User Model by PLSA}. A typical method of this class is proposed in~\cite{wu2008incremental}. This work presents an incremental automatic expert recommendation framework based on PLSA. It considers both users' interests and feedback and takes questions as documents.
It further uses PLSA to model the question to gain its distribution on topics, followed by representing users as the average of topic distributions of all the questions that he accesses to facilitate recommendation.

A most important variant of PLSA is probably the Dual Role Model (DRM) proposed by Xu et al.~\cite{xu2012dual}.
Instead of combining the consideration of a user as a requester and an answerer, DRM separately models users' roles as requesters and as answerers and derive the corresponding probabilistic models based on PLSA.
Depending on the modeling approach of user's role, DRM diverges into \textit{independent DRM}, a type of method modeling user role indirectly, and \textit{dependent DRM}, a method which learns the role model directly.
In particular, the independent DRM assumes all users are independent of each other and models each user individually.
In contrast, dependent DRM considers the dependence between users. Besides modeling users' topic distribution as requesters and answerers, it additionally models the relationship between answerers and requesters for better performance.

\subsubsection{LDA and Its Variants}

The \textit{Latent Dirichlet Allocation (LDA)} model~\cite{blei2003latent} is probably the most widely used topic model among all existing topic models developed.
In LDA, the topic mixture is drawn from a conjugate Dirichlet prior that remains the same for all users.
More specifically, LDA assumes a certain generative process for data. To generate a user profile, LDA assumes that for each user profile a distribution over topics is sampled from a Dirichlet distribution. In the next step, for each word in the user profile, a single topic is chosen according to this topic distribution. Finally, each word is sampled from a multinomial distribution over words specific to the sampled topic.
Here, we briefly review two important classes of LDA variants that have been applied for the expert recommendation in CQA:

\textit{Segmented Topic Model (STM)}~\cite{du2010segmented}. This is a topic model that discovers the hierarchical structure of topics by using the two-parameter Poisson Dirichlet process~\cite{pitman1997two}.
As a four-level probabilistic model, STM contains two levels of topic proportions and shows superiority over traditional models. Instead of grouping all the questions of a user under a single topic distribution, it allows each question to have a different and separate distribution over the topics.
A user profile is considered a document that contains questions (segments). The above distributions cover the expertise set of a user, the topics of each question in the profile, as well as the correlation between each profile and its questions.

\textit{TagLDA}~\cite{sahu2016taglda}. This method uses only tag information to infer users' topical interest. It is more efficiently, but the effectiveness is dependent on the accuracy and availability of tags.

\subsubsection{Expertise-aware LDA}

The work in~\cite{tian2013predicting} considers both the topical interest and expertise of a user relevant to the topics of the given question. It also uses LDA to identify topical interest from previous answers of the user, but additional compute the expertise level of users using collaborative voting mechanism.
Sahu et al.~\cite{sahu2016taglda} incorporate question tags and related voting information in LDA to compute user expertise, where user expertise is computed based on both the topical distribution of users and voting information under the same question tags.

\subsubsection{Other Topic Models}

Besides the famous QLL and LDA models, Zhou et al.~\cite{zhou2009routing} propose a method that groups threads (i.e., a question and related answers) of similar content into clusters to build a cluster-based thread for each user. Each cluster represents a coherent topic and is associated with users to indicate the relevance relationship. The ranking score for a user is then computed based on the aggregation of the relevance of the user to all clusters given a new question.
Guo~\cite{guo2008tapping} proposes a user-centric and category-sensitive generative model for discovering topics, named User-Question-Answer (UQA). The work incorporates topics discovered by UQA model with term-level matching methods to recommend experts and increase the participation rate of users in CQA.
In this model, each user is considered as a pseudo-document which is a combination of all the questions the user has asked and all the answers the user has provided in reply to other users' questions.
More methods can be derived based on this model as well as the combinations of these methods.

\subsection{Network-based Methods}

The network-based methods evaluate users' authoritativeness in a user-user network formed by their asking-answering relations and recommend the most authoritative users as experts for a new question.
The simplest network-based method uses
\textit{Indegree}~\cite{zhang2007expertise} to rank and recommend users.
In particular, an indegree score equals the number of other users a user has helped by answering their questions, represented by an arrow from the requester to the answerer in the user-user network. 
Since frequent posters tend to have a significant interest in the topic and a larger degree of a node usually correlates with answer quality~\cite{jeon2006framework,zhang2007expertise}, this method regards the users with higher degrees as better answerers for the recommendation.
The mainstream of this category include three families of methods based on PageRank~\cite{borodin2005link}, HITS~\cite{kleinberg1999authoritative}, and ExpertiseRank~\cite{zhang2007expertise}, respectively. We will also briefly introduce several other network-based methods to gain a comprehensive view of the related techniques.

\subsubsection{PageRank and Its Variants}

PageRank~\cite{jurczyk2007discovering,jurczyk2007hits} uses nodes to represent users, and a directed edge to indicate one user (i.e., the source node) answers the questions of another user (i.e., the destination node).
It estimates the likelihood that a random walk following links (and occasional random jumps) will visit a node.
Each edge is associated with an affinity weight that measures the times that the answerer has replied to the requesters' questions.
Two users are connected if their affinity weight is greater than 0, and the transition probabilities between nodes are obtained by normalizing the affinity weights.
Now, the algorithm has been extended to bias the jump probability for particular topics~\cite{haveliwala2002topic} and many others static web ranking tasks.
Choetkiertikul et al.~\cite{choetkiertikul2015will} also use PageRank, but they measure the weights differently, i.e., evaluating the number of users' tags and activity times in common as weights between the users who have asking-answering relations.

The main variants of PageRank include SALSA~\cite{lempel2001salsa}, EntityRank~\cite{cheng2007entityrank}, TwitterRank~\cite{weng2010twitterrank}, and AuthorRank~\cite{liu2005co}.
They differ from the above PageRank-based methods in focusing on some specific application domains. However, they still have the potential to be generalized to broader scenarios.

\subsubsection{HITS and Its Variants}
Different from PageRank, which does not distinguish between hub and authority nodes, the HITS algorithm is based on the observation that there are two types of nodes: (1) hubs, which links to authoritative nodes; (2) authorities, which provide useful information on the given topics. HITS assigns each node two scores: hub score and authority score. A hub score represents the quality of outgoing links from the nodes while authority represents the quality of information located on that nodes.
A typical work based on HITS is proposed by Jurczyk et al.~\cite{jurczyk2007discovering,jurczyk2007hits}.
Instead of the times of the answerer replying to the requester's questions, this work models the weights of edges to indicate answer quality in HITS, based on users' explicit feedback, e.g., thumb up/down from users, and whether the answer is the best answer. The results show that the HITS algorithm outperforms the methods based on simple graph measures such as in-degree.

An important variant of HITS is proposed by Shahriari et al.~\cite{shahriari2015community}.
Instead of considering the entire user-user network as a single community, this method regards the network as the combination of multiple user communities.
Thereby, it detects overlapping communities~\cite{shahriari2015disassortative} and differentiate the impact of intra-community and inter-community users to other users in the user-user network.

\subsubsection{Expertise-aware Network-based Methods}

Zhang et al.~\cite{zhang2007expertise} propose ExpertiseRank based on the intuition that an answerer whore replies to a question usually has superior expertise than the requester on the specific topic.
Their experimental results indicate that for a closed domain such as the Java developer forum, ExpertiseRank performs better than general graph-based algorithms like PageRank and HITS.
ExpertiseRank considers not only how many other users a user has helped but also which other users the user has helped.
It propagates the expertise scores of users through the question-answer relationship in a user-user network. The intuition behind ExpertiseRank is that one should get more credit for answering the questions of a user with higher expertise rather than the questions of a user with lower expertise.

The ExpertiseRank computation finally derives a score for each user, called z-score, based on which to quantify their authoritativeness. \textit{Z-score~\cite{zhang2007expertise}} combines users' asking and replying patterns to measure how many standard deviations above or below the expected `random' value a user lies. The more questions a user has answered and the fewer questions the user has asked, the higher the Z-score of this user. Therefore, this method recommends users with the highest z-scores as experts.
The experts obtained by this method should answer many questions and ask very few questions.

Besides ExpertiseRank,
Jurczyk et al.~\cite{jurczyk2007discovering} incorporate users' authoritativeness on a given question (estimated by the tags of users' posts) and users' answer quality (predicted based on their past answer activities); the method in~\cite{bouguessa2008identifying} differs in determining user quality by the number of best answers they provide.
Zhou et al.~\cite{zhou2009routing} use the post content of users to compute user expertise and reply frequencies of users computed by PageRank to re-rank users.
They further use inverted indexes and threshold algorithm~\cite{fagin2003optimal} to store and retrieve pre-computed intermediate scores to accelerate the computation.
The final score of a user takes the product of the results of language model and the results of PageRank.

\subsubsection{Other Network-based Methods}

Here, we review some typical enhancement or extension of the traditional link analysis methods as follows:
Zhu et al.~\cite{zhu2011towards,zhu2014ranking} additionally consider the category relevance of questions and rank user authority in an extended category link graph;
Liu et al.~\cite{liu2011competition} comprehensively utilize multiple types of relationships between requesters and answerers, and between the best answerers and other answerers to find expert users;
rather than leveraging the asking-answering interactions, Lai et al.~\cite{lai2012question} employ the endorsement relationship among users to form user reputation graphs, based on which to recommend experts.
Similarly, Lin et al.~\cite{lin2017smartq}compute user reputation based on their trust relationship in a user-user network. By assuming each question has a known category and theme, they cluster users based on their reputation scores on each question theme and evaluate users based on their theme-specific reputation.
Liu et al.~\cite{liu2013integrating} incorporate user-subject relevance (computed by cosine similarity) and user reputation with users' category-specific authoritativeness obtained from link analysis for the expert recommendation.

The latest network-based method is proposed by Liu et al.~\cite{liu2017knowledge}.
This work routs questions to potential answerers from the viewpoint of knowledge graph embedding and integrates topic representations with network structure into a unified question routing framework.
The framework takes into account various types of relationships among users and questions. It is demonstrated to increase the answer rates of questions by using the recommended experts.

\subsection{Classification Methods}

When regarding experts as a particular class of users among all users, the problem of identifying the experts can be easily transformed into a classification problem that aims to distinguish such a particular class of expert users from the other users.
Compared with the other methods, classification methods can easily apply multiple aspects of features from the user, question, answer, or user-user interaction's perspectives, to the expert recommendation problem.
For example, Pal et al.~\cite{pal2010expert} use three classes of features and train a binary classifier to distinguish experts from ordinary users.  These features include question features (e.g., question length, word n-gram), user features (e.g., number of answers and number of best answers provided by a user), and user feedback on answers (e.g., user votes and comments to the answers), 

Support Vector Machine (SVM) is the most used classification method for distinguishing expert from those non-experts.
Beside diverse methods, the classification methods have used different features besides the basic question and user features, such as part-of-speech features, graph features to trains their models.
For example, Pal et al.~\cite{pal2011early} extract features by modeling users' motivation and ability to help others and use SVM and C4.5 decision tree separately for expert detection.
Zhou et al.~\cite{zhou2012classification} also use SVM but they define both local and global features on questions, user history, and question-user relationship and additional consider KL-divergence as a new feature.
Ji et al.~\cite{ji2013learning} additionally use text similarities as features to train SVM and one of its variant, RankingSVM.

Besides, more methods such as random forests (RF)~\cite{choetkiertikul2015will} and Naive Bayes~\cite{van2015early} are used by existing studies.
Some typical work includes:
Le et al.~\cite{le2016retrieving} consider more new features such as community features (e.g., average score of posts, average number of comments, average favorites marked), temporal features (e.g., time gaps between posts), and consistent features (e.g., scores of the posts, time gap between recent posts).
They also try some new methods like logistic regression and adaptive boosting in addition to decision trees and random forest for the classification.
As an enhancement to decision trees, Dror et al.~\cite{dror2011want} propose a representation model based on multi-channel vector space model, where users and questions are represented as vectors consisting of multi-dimensional features. Then, the matching degree between a user and a question is learned from their respective features using a binary classifier called Gradient Boosted Decision Trees (GBDT).

Instead of the conventional features used for identifying experts, Pal et al.~\cite{pal2010expert} use a different criterion, \textit{question selection bias}, for recommending experts, based on the assumption that experts prefer answering questions to which they bear a higher chance of making a valuable contribution. o
They use the probability of answering questions of different value as the feature vector and employ two methods, logistic regression and Gaussian Mixture Model, to solve the binary classification problem.
In a later version, Pal et al.~\cite{pal2012exploring} use different equations to estimate the value of existing questions and to model the selection probabilities of questions by users.
The method shows better performance of the Bagging metaclassifier over several single-version classification algorithms
The work also partially confirms that experts have some bias on question selection based on the \textit{existing value} of answers to the questions.

Given the advantages of ranked recommendation results over unranked results,
Ji et al.~\cite{ji2013learning} propose RankingSVM, a ranking model based on SVM, for the expert recommendation.
Burel et al.~\cite{burel2015predicting} extract patterns from the question-selection behaviors of users in a Q\&A community and then use Learning to Rank (LTR) models to identify the most relevant question to a user at any given time. They further employ Random Forests, LambdaRank~\cite{burges2007learning}, and ListNet~\cite{cao2007learning} to derive a pointwise method, a pairwise method, and a listwise method, respectively, to gain ranked expert list along with the classification process.
Similarly, Cheng et al.~\cite{cheng2015exploiting} also formalize expert finding as a learning-to-rank task. However, they leverage users' voting information on answers as the ``relevance" labels and utilize LambdaMART to learn ranking models which directly optimizes a rank-based evaluation metric, normalized discounted cumulative gain (nDCG). Logistic regression~\cite{wang2016answerer} has also been used recently to facilitate both ranking and classification.

\subsection{Expertise Probabilistic Models}

Dom et al.~\cite{dom2008bayesian} propose the only probabilistic model that focuses on user expertise for the expert recommendation in CQA. They use a Bayesian probabilistic model to obtain a posterior estimate of user credibility thereby recommending experts for a given question.
This work assumes user expertise conforms to the Bernoulli distribution (or the mixture of two beta distributions) and uses Maximum A Posteriori (MAP) estimation to make predictions.
It then ranks users according to their probabilities of providing the best answer to the given question, characterized by the probability of each user to be awarded the best answer on a question given the user's question-answering history.
The work also discovered that Bayesian smoothing performs better than several other smoothing methods such as maximum a priori estimation, maximum likelihood (ML) estimation, and Laplace smoothing.

\subsection{Collaborative Filtering Methods}

Given the advantages such as flexibility and scalability of collaborative filtering (CF) methods, esp., the matrix factorization techniques~\cite{koren2009matrix} in the recommendation domain, some researchers seek to use CF for the expert recommendation in CQA.
For example, Cho et al.~\cite{cho2015recommending} consider that case that only one expert will be designated to answer each question. They then capture users' behavioral features as potential answerers by matrix factorization~\cite{singh2008relational}.
Regularization terms are incorporated to represent user interest, user similarity, and users' probability to answer for better performance.

Instead of the basic matrix factorization model, Yang et al.~\cite{yang2014tag} employ probabilistic matrix factorization (PMF), which combines generative probabilistic model and matrix factorization, to address the expert recommendation problem. In particular, PMF learns the latent feature space of both users and tags to build a user-tag matrix~\cite{mnih2008probabilistic}, which is then used to recommend experts given a new question.

\subsection{Hybrid Methods}

To comprehensively take into account multiple aspects of clues, some researchers propose hybrid methods that combine different aspects of concerns techniques and different techniques for better recommendation results.
Here, we review the typical hybrid methods as follows.

\subsubsection{Language Model + Topic Model}

Liu et al.~\cite{liu2010predicting} combine QLL and LDA and show the hybrid approach outperforms either of the original models.

\subsubsection{Topic Model + Network-based Method}
Zhou et al.~\cite{zhou2012topical} identify authoritative users by considering both the link structure and topic information about users.
They first apply LDA to obtain user-topic matrix and topic-word matrix, and then use PageRank and Topical PageRank for the expert recommendation.
Liu et al.~\cite{liu2015zhihurank} propose a topic-sensitive probabilistic model to estimate the user authority ranking for each question.
The model is based on PageRank incorporate with topical similarities between users and questions.
A very similar method named topic-sensitive link analysis is proposed by Yang et al.~\cite{yang2016finding}.
Recently, Rao et al.~\cite{rao2016user} propose a similar approach that recommends experts based on users' topical relevance and authoritativeness given a new question.
They also use LDA to discover topics but measure users' authoritativeness for each topic based on the `like' relationship among users in a social network.

Zhao et al.~\cite{zhao2013topic} use the TEL model to generate topic information, based on which to model experts over topics.
TEL is based on LDA and is a unified model combining both graph-based link analysis and content-based semantic analysis for expert modeling in CQA.
Instead of using a fixed user-answering graph to influence the prior of expert modeling, TEL highlights the causal relationship between topics and experts by using both user-user matrix and question-user matrix to represent a user's contribution to another user or a question, with best answers given higher weights.
It is compared with two baseline topic modeling methods, one recommending experts based on requester-answerer interactions and the other recommending experts based on question-answerer interactions, which show its better performance.

Zhou et al.\cite{zhou2012topic} extend PageRank with the topic-sensitive probabilistic model by considering topical similarity.
In particular, the method improves the PageRank-based expert recommendation methods by running PageRank for each topic separately, with each topic-specific PageRank prefers those users with high relevance to the corresponding topic.
As a further step, Yang et al.~\cite{yang2013cqarank} jointly model topics and expertise to find experts with both similar topical preference and superior topical expertise on the given question.
The method integrates textual content model (GMM) and link structure analysis and leverages both tagging and voting information.
It linearly incorporates the estimated user topical expertise score into the recursive PageRank score computation formula and extends PageRank to make the final recommendation.

\subsubsection{Language+Topic+Network-based Model}
Liu et al.~\cite{liu2010predicting} assume that more authoritative answerers may give more accurate answers, and more active answerers may be more willing to answer new questions.
They linearly combine QLL and LDA models to compute relevance, and additionally consider both user activity and authority information for the recommendation.

\subsubsection{Topic Model + Classification Method}

RankingSVM~\cite{ji2013learning} employs LDA to calculate text similarity and use this similarity as a feature in a classification method for the expert recommendation. The experimental results show the resulting method achieves better performance than SVM.

\subsubsection{Topic Model +Collaborative Filtering}

Yan et al.~\cite{yan2012new} combine topic model and tensor factorization for the expert recommendation.
They train an LDA via Gibbs Sampling with a manually defined topic number, followed by performing tensor factorization (TF) based on ``\textit{requester-topic-answerer}" triples via gradient descent to compute the recommendation scores of users.

\subsubsection{Network-based Method + Clustering}

Following the similar idea of geo-social community discovery~\cite{yin2016discovering} in Point of Interest (POI) recommendation,
Bouguessa et al.~\cite{bouguessa2008identifying} incorporate clustering methods with network-based measures for the expert recommendation.
In particular, they consider the number of best answers as an indicator of authoritativeness of a user in a user-user network, where users are connected via directed edge from requesters to best answerer, with the edges weighted by the number of best answers in-between.
In particular, they model the authority scores of users as a mixture of gamma distribution and use the Fuzzy C-Means algorithm to partition users into different numbers of clusters. They further use Bayesian Information Criteria (BIC) to estimate the appropriate number of mixtures. Finally, the users are classified into to classes, one representing authoritative users with high in-degrees and the other non-authoritative users with low in-degrees.
In this way, the method can automatically surface the number of experts in a community rather than producing a ranked list of users.

\section{Comparison and Discussion}
\label{comparison}

To gain a better understanding of state of the art, we first summarize the existing expert recommendation methods concerning the used dataset, the required input \& output, and the evaluation metric.
We further compare and discuss the methods from three perspectives: the covered aspects of concern, reliance on sufficient data, and complexity, to identify their strengths and pitfalls.
The three perspectives reflect the methods' capability in the recommendation power, applicability (robustness to cold start or sparse data), and easiness of usage (implementation difficulty), respectively.

\subsection{Datasets}

In this section, we list the most used datasets by existing expert recommendation research.
These datasets include both the real-world and synthetic ones, as well as those that do not belong to but are readily applicable to evaluating the methods for the expert recommendation in CQA.
Among the real-world datasets, only the first two represent the dominant datasets used by most existing research while all the others are less used.

\subsubsection{Yahoo! Answers}
Yahoo! Answers~\cite{jurczyk2007discovering,guo2008tapping,suryanto2009quality,li2010routing,li2011question,zhou2012classification,yan2012new,zhao2013topic,xu2012dual,zhou2012topical} is perhaps the most popular and most studied datasets in Q\&A related research.
The characteristics of Yahoo! Answers, such as the diversity of questions and answers, the breadth of answering, and the quality of those answers, are first investigated by Adamic et al.~\cite{adamic2008knowledge} in 2008.
In particular, they use \textit{user entropy} to indicate a user's concentration or focus on the different categories of topics.
They further cluster questions and answers based on the content to understand users' activity on different topics in CQA. The results showed that the majority of users participated in a
small number of topics. These features set the practical foundation for predicting answer quality by the amount of work and activities of users.
Since each question has at most one best answer, the amount of ground truth might be sparse when only a part of the entire dataset is used in experiments.
For this reason, some researchers set up their own criteria to determine whether an answer is a ``good" answers or not, to expand the training and test set for their methods.
For example, Li et al.\cite{li2010routing} label an answer a ``good" answer either when it is selected as the best answer or when it obtains more than 50\% of up-votes for all the answers of the question. Meanwhile, one answer is labeled as a ``bad" answer if it receives more than 50\% of rate-downs for all answers of the question.

\subsubsection{Stack Overflow}
Stack Overflow~\cite{riahi2012finding,pal2010expert,pal2012exploring,chang2013routing,du2010segmented,sahu2016taglda,le2016retrieving,choetkiertikul2015will,cheng2015exploiting,yeniterzi2015moving} involves over five million users and
content about 11,053,469 questions, among which only 73\% have
received answers and closed and 55\%, i.e., over six million questions,
have accepted best answers (as of 10 March 2016).
Like the Yahoo! Answers dataset, the records in the Stack Overflow dataset is massive, and most existing research sample a subset of the entire dataset for study.
For example, Pal et al.~\cite{pal2012exploring} sample a small dataset of 100 users and employ two expert coders to label the 100 users as either experts or non-experts.
It turns out that the inter-rater agreement between the expert coders is 0.72 (Fleiss kappa with $95\% CI$, $p \sim 0$), indicating the high agreement between the raters is not accidental. Out of the 100 users, 22 are labeled as experts and rest as non-experts.

\subsubsection{Other CQA Datasets}
\textit{TurboTax Live Community (TurboTax)}\footnote{http://ttlc.intuit.com/, May 2018.}~\cite{pal2010expert,pal2012exploring,pal2011early}: this is a Q\&A service related to preparation of tax returns.
TurboTax has employees that manually evaluate an expert candidate on factors, such as correctness and completeness of answers, politeness in responses, language and choice of words used. They also have some labeled experts.

\textit{Quara}~\cite{zhao2015expert,zhao2015cold}: a general and probably the world' largest Q\& A website that covers various topics.

\textit{Java Developer Forum}~\cite{zhang2007expertise}: an online community where people
come to ask questions about Java. It has 87 sub-forums that focus on various topics concerning Java programming. There is a broad diversity of users, ranging from students learning Java to the top Java experts. A typical sub-forum, e.g., ``Java Forum", a place for people to ask general Java programming questions, has a total of 333,314 messages in 49,888 threads as of as early as 2007.

\textit{Naver KnowledgeCiN}: the largest question-answering online community in South Korea. Nam et al.~\cite{nam2009questions} analyze the characteristics of knowledge generation and user participation behavior in this website and finds that altruism, learning, and competency are often the motivations for top answerers to participate.

\textit{Baidu Knows}\footnote{http://zhidao.baidu.com/, May 2018.}: a Chinese language CQA service, where a member can put questions with bounty to promote others answering it. Once the answer is accepted, it turns into search result of relevant questions.

\textit{Tripadvisor forums}\footnote{http://www.tripadvisor.com/, May 2018.}~\cite{zhou2009routing}: a travel-related websites with user-generated content focusing on accommodation bookings. The service is free to users, who provide feedback and reviews to hotels, accommodation facilities, and other traveling related issues.

\textit{Sogou Wenwen}~\cite{yan2012new}: formerly known as Tencent Wenwen or Soso Wenwen, is similar to Quora and also run with credit points and reputation points. Users can obtain points by asking or answering questions and use them as bounty.

\textit{Iask}\footnote{http://iask.sina.com.cn/, May 2018.}~\cite{liu2010predicting}: a leading web 2.0 site in China. The working mechanism is similar to Baidu Knows, while in Iask, a requester can increase bounty to extend 15 days before question closed due to a previously accepted answer.

\textit{Other datasets on Stack Exchange}: such as computer science\footnote{http://cs.stackexchange.com/, May 2018.}, fitness\footnote{http://fitness.stackexchange.com/, May 2018.}~\cite{shahriari2015community}, and cooking\footnote{http://cooking.stackexchange.com/, May 2018.}. There are total 133 communities for knowledge sharing and question answering, covering enormous topics on Stack Exchange.

\textit{Estonian Nature forum}~\cite{shahriari2015community}: a Q\&A website popular in Estonia.

\textit{MedHelp}\footnote{http://www.medhelp.org/, May 2018.}~\cite{cho2015recommending}: a website which partners with doctors from hospitals and research facilities to provide online discussion and to satisfy users' medical information needs.

\subsubsection{Synthetic Dataset}

Generally, no single method outperforms all the others on all the datasets for two main reasons:
first, online communities usually have different structural characteristics and lead to differences in the performance of methods~\cite{zhang2007expertise}; second, the same users may behave differently in different communities due to various reasons such as the subjectivity and rewarding mechanism of a Q\&A system.
Given the lack of benchmarks to evaluate the different methods, it has become a common practice to conduct controlled experiments with simulated datasets to test how a method performs under different scenarios.
We will not give more introduction to the synthetic datasets due to the significant variances in the assumptions and conditions to generating these datasets.

\subsubsection{Non-CQA Datasets}

There are plenty of datasets do not belong to the Q\&A domain but are readily applicable to or have been used for the study of expert recommendation methods for the CQA.
A slight difference of the methods developed based on studying these datasets is that they most
aim to rank and find the most best-skilled or authoritative users given an existing domain or topic instead of a new question.
These datasets include \textit{co-authorship network}~\cite{liu2005co,li2009searching,daud2013finding} such as DBLP~\cite{deng2008formal,hashemi2013expertise,mimno2007expertise}, \textit{social networks}~\cite{horowitz2010anatomy, bagdouri2015cross,richardson2011supporting}, \textit{microblogs}~\cite{java2006modeling,kempe2003maximizing,pal2011identifying} such as Twitter~\cite{weng2010twitterrank}, \textit{Email network}~\cite{campbell2003expertise,dom2003graph,shetty2005discovering}, Internet forums~\cite{zhang2007expertise}, log data~\cite{mockus2002expertise}, e-Learning platform~\cite{wei2007measuring}, Usenet newsgroups~\cite{fisher2006you,viegas2004newsgroup}, Google Groups~\cite{welser2007visualizing}, general documents~\cite{fu2007cdd}, and enterprise documents~\cite{balog2006formal,balog2007broad,petkova2008hierarchical} such as Enterprise track of TREC~\cite{pasca2001high,fang2007probabilistic,macdonald2006voting}.

\subsection{Input and Output}
To ease illustration, we first summarize the typical inputs and outputs of existing expert recommendation methods in
Table~\ref{Table2}.
Here, we list five categories of commonly used inputs for expert recommendation methods. These inputs are either textual, numerical, or relational information, while the outputs, i.e., the recommended experts, are either ranked or unranked, depending on the methods adopted.

\tabcolsep 9pt
\begin{table*}[!htb]
\centering
\caption{\label{Table2} Typical Inputs and Outputs of Expert Recommendation Methods}\vspace{-2mm}
{\footnotesize
\resizebox{\linewidth}{!}{
\begin{tabular*}{\linewidth}{c|c|c|l|l}
\hline\hline\hline
Type & Category &  Id & Input/output name & Input/output type\\
\hline
\multirow{13}{*}{Input} & Question profile
 & I0
   & content (and category) of the given question & textual\\ \cline{2-5}

 & \multirow{4}{*}{User profile}
 & I1
   & users' question history & user-question mapping\\ 
 & & I2
 &  users' answer history & user-answer mapping  \\ 
 & & I3
 &  users' historical viewing and answering activity & multiple user-question mapping\\ 
 & & I4
 &  timestamps of users' answering activity & numerical \\ \cline{2-5}
 
& \multirow{5}{*}{\tabincell{c}{Historical\\[-5pt] questions \& answers}}
  & I5
 &  question content  & textual  \\ 
 & & I6
 &  question category info & textual  \\ 
  & & I7
 &  question tags & textual \\ 
 & & I8
 &  answer content & textual\\ 
 & & I9
   & best answer info & answer--\{0,1\} mapping\\  \cline{2-5}
 
 & \multirow{2}{*}{Social profile}
    & IA
   & voting info & numerical \\ 
 
 & & IB
 &  user reputation & numerical \\  \cline{2-5}
 
& Network profile
 & IC
   & question-answer relations among users & directed user-user mapping \\ \hline
 
\multirow{2}{*}{Output} & \multirow{2}{*}{ \tabincell{c}{Recommended\\[-5pt] experts}}
& O1
   & an unranked group of experts & set \\ 
   
 & & O2
   & a ranked list of experts & list \\
\hline\hline\hline
\end{tabular*}
}}
\end{table*}
\baselineskip=13.2pt plus.2pt minus.2pt
\parskip=0pt plus.2pt minus0.2pt

Based on the input/output list, we further present a comparison of the representative methods with respect to their inputs and outputs in Table~\ref{Table3}.
Some methods may use derived features from the original inputs as additional inputs. For example, a classification method may use the length of questions (implied by question content), total question number of users (implied by users' question history), and total answer number of users (implied by users' answer history) as additional features to train their models.

\tabcolsep 3.8pt
\begin{table*}[!htb]
\centering
\caption{\label{Table3} A Comparison of Inputs and Outputs of Representative Expert Recommendation Methods}\vspace{-2mm}
{\footnotesize
\noindent\makebox[\linewidth]{%
\begin{tabular*}{\linewidth}%
{c|c|ccccccccccccccc}
\hline\hline\hline
Category & Representative method & I0&I1&I2&I3&I4&I5&I6&I7&I8&I9&IA&IB&IC&O1&O2\\
\hline
\multirow{3}{*}{\tabincell{c}{Simple\\[-5pt] methods}}
 & Votes
   &&&$\surd$&&&&&&&&$\surd$ &&&&$\surd$  \\ 
   
 & Best answer proportion
 &  && $\surd$&&&&&&&$\surd$&&&&&$\surd$ \\ 

& Consine similarity based on TF-IDF~\cite{baeza1999modern}
&$\surd$&&$\surd$&&&$\surd$&&&&&&&&&$\surd$
\\ \hline

\multirow{3}{*}{\tabincell{c}{Language\\[-5pt] models}}
& QLL~\cite{liu2005finding,lavrenko2001relevance,liu2004cluster}
 & $\surd$ && $\surd$&&&$\surd$&&&&&&&&&$\surd$ \\
   & Category-sensitive QLL~\cite{li2011question,li2011question}
& $\surd$ && $\surd$&&&$\surd$&$\surd$&&&&&&&&$\surd$  \\
  & Expertise-aware QLL~\cite{zheng2012algorithm,li2010routing}
& $\surd$ && $\surd$&&&$\surd$&&&&$\surd$&&&&&$\surd$
  \\ \hline
  
\multirow{2}{*}{\tabincell{c}{Topic\\[-5pt] models}}
 & PLSA~\cite{qu2009probabilistic,wu2008incremental}, LDA~\cite{blei2003latent}, STM~\cite{du2010segmented}, UQA~\cite{guo2008tapping}
   & $\surd$ && $\surd$&&&$\surd$&&&&&&&&&$\surd$  \\
 & DRM~\cite{xu2012dual}
   & $\surd$ &$\surd$& $\surd$&&&$\surd$&&&&&&&&&$\surd$  \\
 & TagLDA~\cite{sahu2016taglda}
 & $\surd$ && $\surd$&&&&&$\surd$&&&&&&&$\surd$ \\ \hline

\multirow{5}{*}{\tabincell{c}{\tabincell{c}{Network-based\\[-5pt] methods}}}
 & Indegree~\cite{zhang2007expertise}, PageRank~\cite{choetkiertikul2015will,haveliwala2002topic}, HITS~\cite{jurczyk2007hits,shahriari2015community}
    &&&&&&&&&&&&&$\surd$&&$\surd$ \\
 & z-score~\cite{zhang2007expertise}, Expertise-aware methods~\cite{zhang2007expertise,jurczyk2007discovering,bouguessa2008identifying,zhou2009routing,fagin2003optimal}
&&&$\surd$&&&&&&&$\surd$&&&$\surd$&&$\surd$\\
 & Reputation-aware methods~\cite{lai2012question,lin2017smartq}
 &&&&&&&&&&&&$\surd$&$\surd$&&$\surd$ \\
 & Category-sensitive methods~\cite{zhu2011towards,zhu2014ranking}
 &$\surd$&&&&&&$\surd$&&&&&&$\surd$&&$\surd$  \\
 & Graph embedding method~\cite{liu2017knowledge}
   &$\surd$&$\surd$&$\surd$&&&$\surd$&&&$\surd$&$\surd$&&&$\surd$&&$\surd$  \\\hline
\multirow{2}{*}{\tabincell{c}{Classification\\[-5pt] methods}}
   & \tabincell{c}{SVM~\cite{pal2011early,zhou2012classification}, C4.5~\cite{pal2011early}, RF~\cite{choetkiertikul2015will}, GBDT~\cite{dror2011want}}
   &$\surd$&&$\surd$&$\surd$&&$\surd$&&$\surd$&&$\surd$&$\surd$&$\surd$&$\surd$&$\surd$&
   \\
 & LTR~\cite{burel2015predicting}
 
   &$\surd$&&$\surd$&$\surd$&&$\surd$&&$\surd$&&$\surd$&$\surd$&$\surd$&$\surd$&&$\surd$
    \\\hline 
 
 \multirow{2}{*}{\tabincell{c}{Expertise probability\\[-5pt] model}}
   & \multirow{2}{*}{Bernoulli MAP model~\cite{dom2008bayesian}}
    &  && \multirow{2}{*}{$\surd$}&&&&&&&\multirow{2}{*}{$\surd$}&&&&&\multirow{2}{*}{$\surd$}
    \\
    &&&&&&&&&&&&&&&& \\\hline 
\multirow{2}{*}{\tabincell{c}{Collaborative\\[-5pt] Filtering}}
   & MF~\cite{cho2015recommending}
   & $\surd$ &&$\surd$ &$\surd$&&$\surd$&&&&&&&&&$\surd$
   \\
 & Tag-based PMF~\cite{yang2014tag}
& $\surd$ &&$\surd$ &$\surd$&&&&$\surd$&&$\surd$&&&&&$\surd$
    \\\hline 
    
\multirow{4}{*}{\tabincell{c}{Hybrid\\[-5pt] methods}}
   & \tabincell{c}{QLL+LDA~\cite{liu2010predicting}, Topical PageRank~\cite{zhou2012topical},\\[-3pt] TEL~\cite{zhao2013topic},LDA+TF~\cite{yan2012new}}
   & $\surd$ && $\surd$&&&$\surd$&&&&&&&&&$\surd$  
   \\
 & Topical PageRank+Expertise~\cite{yang2013cqarank}
 & $\surd$ && $\surd$&&&$\surd$&&&&$\surd$&&&&&$\surd$  
   \\
 & QLL+LDA+userActivity+Indegree~\cite{liu2010predicting}
 & $\surd$ && $\surd$&&$\surd$&$\surd$&&&&$\surd$&&&$\surd$ &&$\surd$  
   \\
 & Indegree+Clustering~\cite{bouguessa2008identifying}
 & $\surd$ && $\surd$&&&$\surd$&&&&$\surd$&&&$\surd$ &$\surd$&  
    \\
\hline\hline\hline
\end{tabular*}
}}
\end{table*}
\baselineskip=13.2pt plus.2pt minus.2pt
\parskip=0pt plus.2pt minus0.2pt

\subsection{Evaluation Metrics}
We summarize three categories of metrics used to evaluate expert recommendation methods for CQA, namely the basic, rank-based, and human-judgment-based metrics. The following subsections introduce the metrics of each category, respectively, where each metric is computed as (the mean of) the average of the metric values over a set of query questions or topics~\cite{li2011question,yan2012new,zhou2009routing,zhou2012topical,shahriari2015community,cho2015recommending}.

\subsubsection{Basic Metrics}

There are four set-based metrics to evaluate an expert recommendation method:

\textit{Precision}~\cite{pal2010expert,pal2011early,zhou2012classification,liu2008predicting}: the fraction of users who are true experts to the given questions, among all the users recommended by a method.

\textit{Recall}~\cite{pal2010expert,pal2011early,zhou2012classification,liu2008predicting,shahriari2015community}: the fraction of users who are recommended by a method and meanwhile turn out to be the real experts, among all the real experts to the given questions.

\textit{F$_1$}-score~\cite{pal2010expert,pal2011early,zhou2012classification,liu2008predicting,le2016retrieving,van2015early}: the harmonious average of the Precision and Recall.

\textit{Accuracy}~\cite{zhou2012classification,liu2008predicting,dror2011want,le2016retrieving}: the fraction of users who are correctly identified as either an expert or an non-expert by a method. The metric integrates the precision of the method in identifying the experts and non-experts.

\subsubsection{Rank-based Metrics}

\textit{Precision at top n (P@n)}~\cite{li2011question,zhou2009routing,yan2012new,chang2013routing,zhao2013topic,srba2015utilizing,zhou2012topical,cho2015recommending,suryanto2009quality}: the percentage of the top-N candidate answers retrieved that are correct.
It is also known as \textit{Precision at top n (P@n)}~\cite{yan2012new,chang2013routing,zhao2013topic,srba2015utilizing} or \textit{Success at top N (S@N)}~\cite{du2010segmented}.
A special case is Precision@1~\cite{zhao2015expert,zhao2015cold} when $n=1$.

\textit{Recall at top N (R@N)}~\cite{yan2012new,chang2013routing,zhao2015cold,suryanto2009quality}, a natural expansion of the basic recall to rank-based scenario, similar to P@n.

\textit{Accuracy by Rank}~\cite{zhao2015cold}: the ranking percentage of the best answerer among all answers.
A similar metric using the best answerer's rank is proposed in~\cite{qu2009probabilistic} and~\cite{xu2012dual}.

\textit{Mean Reciprocal Rank (MRR)}~\cite{li2010routing,li2011question,yan2012new,zhou2009routing,zhou2012topical,liu2015zhihurank,shahriari2015community,yeniterzi2015moving,burel2015predicting}:  the mean of the reciprocal ranks of the first correct experts over a set of questions, measuring gives us an idea of how far down we must look in a ranked list in order to find a correct answer.

\textit{Matching Set Count (MSC) @n}~\cite{chang2013routing,yeniterzi2015moving}: the average number of the questions that were replied by any user ranked within
top $n$ recommended users.

\textit{Normalized Discounted Cumulative Gain (nDCG)}~\cite{zhao2015expert,shahriari2015community}: a number between 0 and 1, measuring the performance of a recommendation system based on the graded relevance of the recommended items.
A variant is \textit{$nDCG@k$}, the division of the raw DCG by the ideal DCG, where $k$ is the maximum number of items to be recommended.

\textit{Pearson Correlation Coefficient}~\cite{jurczyk2007discovering,fagin2003comparing,herlocker2004evaluating}: the correlation degree between the estimated ranking with the ranks of users according to the scores derived from the user feedback.

\textit{Area Under ROC Curve (AUC)}~\cite{dror2011want}: the probability that an expert is scored higher than a non-expert.

\subsubsection{Human Judgment-based Metrics}

\textit{Correctness percentage}: human judgment is necessary in the case where the ground truth is unavailable or hard to be determined automatically.
In such cases, humans usually give either \textit{Yes/No answers}~\cite{guo2008tapping} or \textit{ratings}~\cite{zhang2007expertise} to the recommended users. Then, the system calculates the percentage of correctly recommended users by investigating the agreement between the judgments made by different people~\cite{liu2008predicting,zhou2012topical,bouguessa2008identifying}.

\subsection{Covered Aspects of Concern}

To study the covered aspects of concern of different methods, we summarize the main aspects of concern and their indicators in an expert recommendation problem in Table~\ref{Table4}.
The inputs taken by each method directly reflect the method's covered aspects of concern.
For example, to take user expertise into account, a method needs to consider at least one of the three aspects: user reputation, voting information to answers, and best answer flags of answers. 
An indicator either belongs to or originates from the inputs. It falls into at least one of the two aspects: answer probability and expertise level of users.

\tabcolsep 12pt
\renewcommand\arraystretch{1.3}
\begin{table*}[!htb]
\centering
\caption{\label{Table4} Aspects of Concern and Their Indicators Covered by Representative Expert Recommendation Methods}\vspace{-2mm}
{\footnotesize
\noindent\makebox[\linewidth]{%
\begin{tabular*}{\linewidth}{c|c|c|c}
\hline\hline\hline
Category & Representative method & Answer probability & Expertise level\\
\hline
\multirow{3}{*}{\tabincell{c}{Simple\\[-5pt] methods}}
 & Votes & & vote counts of answers  \\ 
 & Best answer proportion & & best answer ratio \\ 
& Consine similarity of TF-IDF~\cite{baeza1999modern}
& textual relevance & 
\\ \hline

\multirow{3}{*}{\tabincell{c}{Language\\[-5pt] models}}
& QLL~\cite{liu2005finding,lavrenko2001relevance,liu2004cluster}
& textual relevance &  \\ 
   & Category-sensitive QLL~\cite{li2011question,li2011question}
& \tabincell{c}{question category\\[-5pt] textual relevance} & \\ 
  & Expertise-aware QLL~\cite{zheng2012algorithm,li2010routing}
& textual relevance & best answer ratio 
  \\ \hline
 
\multirow{2}{*}{\tabincell{c}{Topic\\[-5pt] models}}
 
 & PLSA~\cite{qu2009probabilistic,wu2008incremental}, LDA~\cite{blei2003latent}, STM~\cite{du2010segmented}, UQA~\cite{guo2008tapping}
& topical relevance &  \\ 
  
 & DRM~\cite{xu2012dual}
& topical relevance &   \\ 
  
 & TagLDA~\cite{sahu2016taglda}
& topical relevance of tags &   \\ \hline

\multirow{7}{*}{\tabincell{c}{\tabincell{c}{Network-based\\[-5pt] methods}}}
 & Indegree~\cite{zhang2007expertise} & a\# of users &  \\ 
 & PageRank~\cite{choetkiertikul2015will,haveliwala2002topic}, HITS~\cite{jurczyk2007hits,shahriari2015community}
& a\# \& q\# of users & a\# \& q\# of users\\

 & z-score~\cite{zhang2007expertise} & a\# \& q\# of users& \tabincell{c}{a\# \& q\# of users,\\[-5pt] best answer number} \\ 
 
 & Expertise-aware methods~\cite{zhang2007expertise,jurczyk2007discovering,bouguessa2008identifying,zhou2009routing,fagin2003optimal}
& a\# \& q\# of users & \tabincell{c}{a\# \& q\# of users,\\[-5pt] best answer number} \\ 
 
 & Reputation-aware methods~\cite{lai2012question,lin2017smartq}
& a\# \& q\# of users & \tabincell{c}{a\# \& q\# of users,\\[-5pt] user reputation} \\ 
 
 & Category-sensitive methods~\cite{zhu2011towards,zhu2014ranking}
& \tabincell{c}{a\# \& q\# of users,\\[-5pt] category relevance} &  a\# \& q\# of users  \\ 
 
 & Graph embedding method~\cite{liu2017knowledge}
& \tabincell{c}{a\# \& q\# of users,\\[-5pt] textual relevance} & \tabincell{c}{a\# \& q\# of users,\\[-5pt] best answer number} \\\hline
 
\tabincell{c}{Classification\\[-5pt] methods}
   & \tabincell{c}{SVM~\cite{pal2011early,zhou2012classification}, C4.5~\cite{pal2011early}, RF~\cite{choetkiertikul2015will},\\[-3pt] GBDT~\cite{dror2011want}, LTR~\cite{burel2015predicting}}
& \tabincell{c}{textual relevance,\\[-5pt] question features,\\[-5pt] user features (e.g., a\#),\\[-5pt] metrics (e.g., z-score)} & best answer number
   \\ \hline
 
 \multirow{2}{*}{\tabincell{c}{Expertise prob.\\[-5pt] model}}
   & \multirow{2}{*}{Bernoulli MAP model~\cite{dom2008bayesian}}
& \multirow{2}{*}{} & \multirow{2}{*}{best answer ratio}
    \\
   & & &  \\\hline 
    
\multirow{2}{*}{\tabincell{c}{Collaborative\\[-5pt] Filtering}}
   & MF~\cite{cho2015recommending}
& textual relevance &
   \\ 
 
 & Tag-based PMF~\cite{yang2014tag}
& textual relevance of tags & best answer ratio
    \\\hline 

\multirow{6}{*}{\tabincell{c}{Hybrid\\[-5pt] methods}}
   & QLL+LDA~\cite{liu2010predicting}, TEL~\cite{zhao2013topic}
& textual \& topical relevance & 
   \\ 
    
 & LDA+TF~\cite{yan2012new}
& textual \& topical relevance & 
   \\ 
   
 & Indegree+Clustering~\cite{bouguessa2008identifying}
& a\# \& q\# of users & best answer number
    \\
   
   & Topical PageRank~\cite{zhou2012topical}
& \tabincell{c}{topical relevance,\\[-5pt] a\# \& q\# of users} & a\# \& q\# of users
   \\ 
   
  &  Topical PageRank+Expertise~\cite{yang2013cqarank}
& \tabincell{c}{topical relevance,\\[-5pt] a\# \& q\# of users} & \tabincell{c}{a\# \& q\# of users,\\[-5pt] best answer number}
   \\ 
 
 & QLL+LDA+userActivity+Indegree~\cite{liu2010predicting}
& \tabincell{c}{textual/topical relevance,\\[-5pt] a\#, q\#, active time of users} & best answer number 
   \\ 
\hline\hline\hline
\end{tabular*}
}
\\\vspace{1mm}\parbox{\linewidth}{Note: q\# and a\# denote the number of questions asked and answered by a user, respectively.}
}
\end{table*}
\baselineskip=13.2pt plus.2pt minus.2pt
\parskip=0pt plus.2pt minus0.2pt

The inputs and outputs only give some intuitive clues of how powerful each method might be.
In fact, it is not only the inputs but also the ways of using these inputs and the underlying techniques that determine a method's capability in adequately addressing the expert recommendation problem.
In the following, we elaborate each aspect of concern and make a comparative discussion of the investigated methods according to their covered aspects of concern.

\subsubsection{Answer Probability}

A large body of the research on the expert recommendation in CQA focuses merely on ranking the relevance of the given question with users' previously answered questions.
The underlying assumption is that users are more likely to answer those questions that are similar to their previously answered ones.
This type covers all the methods that consider users' answer probability yet not users' expertise level in Table~\ref{Table4} and range from the simple Cosine similarity method to QLL, LDA, MF, and further to hybrid methods that combine the above techniques like TEL.

Similar to the general recommendation approach, it is reasonable to assume a user prefers to answer the questions that are similar to her already-answered questions. On the other hand, there is no evidence to show that the stronger relevance of a user to a given question also indicates a higher expertise level of the user in answering that question. Therefore, such methods may not be able to distinguish users' different levels of expertise in either a question or a topic, and the recommended experts may not provide quality answers to the given question.

Other issues with this type of methods are related to the consideration of categories or topics.
While considering the category or topic information enables a method to provide personalized recommendations customized to the given question, it raises additional issues to the method.
First, the category-sensitive methods highly rely on the availability and proper definition of categories. For example, suppose a user has answered a lot of questions about `c++' covered by the category of `Programming,' and is deemed a potential answerer for questions of this category. Given a question related to another programming language like `Python,' which is also covered by the category of `Programming,' recommending this user as an expert may not be appropriate as the user may not be familiar with `Python' as well. The topic-sensitive approach is more reasonable as topics are usually not predefined but dynamically constructed by algorithms, and therefore they can adapt to the ever increasing questions and answers in a Q\&A community.

Second, an inevitable issue with considering categories or topics is that a user may have an expertise in multiple categories or topics.
For category-sensitive methods, although the correlation among categories can be incorporated explicitly, it could be difficult to designate a question to a single category when the question is related to multiple categories.
For the topic-sensitive methods, they mostly discover topics based on probabilistic models such as LDA.
Since the probabilistic models distribute the total probability of 1 among all the topics for each user, having a higher probability on one topic will discount the probability on other topics. However, the fact is, a user could be more relevant to multiple topics than another user simultaneously. This situation has not been sufficiently taken into account by the existing research.

\subsubsection{User Expertise}

There are a few methods that take into account user expertise while neglecting to consider the answer probability of users in the expert recommendation in CQA.
These methods typically include simple techniques (e.g., votes, best answer proportion) and the expertise probabilistic model in Table~\ref{Table4}.
A limitation of these methods is that they only consider the probability of a user giving the best answer under the condition that the user has decided to answer the given question.
The fact is, a user with the appropriate expertise may not answer the question in the first place due to various reasons such as lack of interest or unavailability.
Therefore, as far as answer probability is concerned, the recommended experts may not have a good chance of giving a quality answer.

Another issue with the methods of this type is that they commonly compute a global expertise score for each user while neglecting the fact that user expertise may also be topic-sensitive, similar to answer probability.
Consequently, the recommendation results independent of to the specific problem scenario: given a new question, the recommended experts may perform generally well but unfortunately perform poorly on the given question.
The global expertise model is most suitable for scenarios where the same topic covers all questions.

\subsubsection{Both Aspects}

Given the importance of both aspects of concern, many existing methods, especially the latest ones, combine the two above aspects for the better expert recommendation.
These methods typically include expertise-aware QLL models, all variants of PageRank/HITS, classification methods, collaborative filtering methods, and hybrid methods that combine two or more of the above methods.
The straightforward way of integrating the two aspects is to compute a score for each aspect separately and then combine the two scores into a weighted sum, which would be used as the criterion for ranking users and deriving the recommendation results.
The other methods, including network-based methods, classification methods, and collaborative filtering methods, combine the two aspects of consideration more naturally.

For example, the network-based methods naturally incorporate the two aspects to compute a single criterion, user authoritativeness, for the expert recommendation---the users who have provided many questions yet asked few questions are considered authoritative; such users are generally believed to have both higher probabilities to answer and better expertise in a user-user network. An alternative is to use the related user number to replace question number in the above computation. This replacement slight changes the computation method but does not affect the authoritativeness definition.
Instead of using a single metric such as authoritativeness to recommend experts, the classification methods directly take various factors as features, without explicitly distinguishing the two aspects, to train a classification model.
A limitation with classification methods is that they generally deliver a set of users as experts without further differentiating them.
Therefore, they lack the flexibility to decide how many users to recommend as experts on the fly.

To explicitly utilize the best answer information in a network-based method, a simple approach is to replace the `requester-answerer' relationship in a user-user network into the `requester-best answerer' relationship among users. In this case, a link is drawn from one user to another user only when the second user has provided the best answer to the first users' questions.
The indegrees and other metrics (e.g., authoritativeness) of users derived from the modified model can directly be used to recommend users who are both active and have the right expertise. In this way, the recommended experts are those who have provided best answers to the largest numbers of other users and have been answered by the fewest other users on their raised questions.

Currently, almost all the hybrid methods that cover both aspects involves network-based methods as a component.
These methods therefore still share some drawbacks of the basic network-based methods.
First, the experts recommended by such methods are specific to a user-user network rather than a particular topic or a question.
Intuitively, both the transitivity of users' authoritativeness and the effectiveness of network-based methods depend on the condition that the interactions between users concern only one topic.
Second, although the link structure can, to some extent, implies the correlation among users and questions, the user-user network is not directly related to a user's topical preference. To recommended experts for a given problem, they still need some approach to connect the question to users or their historical questions to make a personalized recommendation. The more recent expertise-aware network-based methods often hybridize with relevance models (e.g., language models and topic models) to overcome the above deficiencies.

Another possible issue with the hybridization of network-based methods with other methods is that authoritativeness is a vague concept, which already, to some extent, implies the interest and expertise of users. Therefore, the combination of techniques may cause the hybrid methods to consider multiple times of the same aspects. The rationale and effect of such hybridization are yet to be examined.

\subsection{Reliance on Sufficient Data}

The cold start or data sparsity problem concerns both the ground truth or numbers of users' activity records, and it turns out to be a common challenge for all the investigated expert recommendation methods.
For the amount of ground truth, the rule of thumb is that the more straightforward methods tend to be less affected by the small ground truth size, as the more complicated methods usually require a larger train set.
For example, the classification methods generally perform better under high-dimensional features given sufficient training data. When the training data is limited, these methods need to restrain the dimensionality to avoid over-fitting. In contrast,  the voting-base techniques require no training and thus unaffected by the size of training data.

The effect of the historical record numbers of users on the recommendation methods is closely related to the early detection of experts, i.e., promoting the participation rate of new experts or rising stars, i.e., the low profile users who have strong potential to contribute to the community later after short observations in CQA~\cite{le2016retrieving}.
The lack of sufficient information in the user profile is, in fact, the primary obstacle towards identifying such early-career experts.
Previous studies~\cite{zhou2012joint} show that only 15.67\% of all users in Yahoo! Answers answered more than four questions. This observation indicates that all these approaches involve only a small portion of highly active users and therefore cannot recommend new questions to the rest of the community. 
If the recommendation method can also involve the good users with few activity records, these users can become motivated to take more intensive participation in a community or even develop into highly active experts.

Despite the significance of early expert detection issue, the markers that reflect the expertise of an ordinary user (e.g., number of answers and number of best answers) are not that strong for a newly joined user. Therefore, not much prior work has researched on finding potential experts in early-stage in CQA~\cite{pal2010expert,pal2012exploring,pal2011early}.
Many existing methods bypass the cold start and data sparsity issues due to their reliance on sufficient data from CQA systems. For example, some approaches consider only those users who previously provided more than 5~\cite{tian2013predicting}, 10~\cite{guo2008tapping} or even 20 answers ~\cite{fang2012question}.
Other methods take only users with a significant number of best answers into consideration (e.g., more than 10~\cite{liu2010predicting} or 20 best answers~\cite{riahi2012finding}).

Among the existing methods, we identify two promising categories of methods that can potentially better detect experts early. One is \textit{semi-supervised Learning methods} (e.g.,~\cite{van2015early}), which regard users who provide above average-best-answers on a topic tag as topical experts. They apply a data-driven approach to predict whether a user will become an expert in the long term. The other is \textit{expertise propagation methods} (e.g.,~\cite{sung2013booming}), which infer or consolidate the expertise of low-profile users by propagating the expertise of old users through their shared activities in a community.

\subsection{Method Complexity}

Generally, the more aspects considered, the better a method can potentially perform, and the more complicated the method could be.
The expertise-aware techniques based on QLL usually combine the two aspects linearly using the simple weight sum method. The primary issue with these methods is the difficulty in allocating the weights wisely among the two aspects.
Usually, they need to resort to human experience or repeated trials in real applications to determine the optimal weights.

Though applicable to the expert recommendation problem, recommendation techniques face severe challenges besides the fundamental issues like the cold start problem.
For example, considering multiple aspects of concerns could make a recommendation technique complex and challenging to optimize.
More recently recommendation methods such as factorization machines may help resolve the problem but have not yet been applied to the expert recommendation in CQA.

Despite the ability to incorporate multiple aspects of concern, there is a lack of universal principle regarding which features to use for the classification methods.
Consequently, the performance of classification methods largely depends on the features used and whether the technique and features fit the size of the labeled data.

\section{Future Directions}
\label{Directions}

After reviewing the state of the art methods, we identify several challenging yet promising issues for the future research. We summarize them as realistic user modeling, recommending experts as a collaborative group, coping with dynamicity, utilization of external data, and comprehensive expert recommendation solutions.
In the following, we review the limited related studies to the above challenges, highlight the significance of and new opportunities for addressing these challenges, and finally, outlook the promising directions for future research.
We hope this discussion could provide novel viewpoints to the existing studies and encourage more future contributions to this promising field of research.

\subsection{Realistic User Modeling}
\label{Aspects}

Expert recommendation relies on effective user modeling.
Intuitively, there exist three aspects of concerns that affect whether a user gives a high-quality answer to a question in a real Q\&A scenario as follows:

\textit{The chance of a user noticing the question}. Since a user may not have an opportunity to see a question, the user may not be an answerer to this question even though the user is an expert. The expert recommendation problem in CQA, however, is based on a different assumption from the real-world scenarios, i.e., how likely a user would answer a question and meanwhile provide a high-quality answer to the question if the user is invited to answer the question. Due to the above difference, when using the real-world labeled data to train the recommendation models, the recommendation methods should better take into account the possibility that a user may not have answered a question just because the user does not have the chance to notice the question. The likelihood that a user would see a question in real-world scenarios depends on various factors such as user availability (e.g., how often a user is online and available to answer questions), user behaviors (e.g., whether the user looks for new questions to answer actively), and other users' activities related to the question (e.g., how widespread the question is among users).

\textit{User's willingness to answer the question}. Even if a user has noticed a question, the user may choose not to answer it. A user's willingness to answer a question also depends on various factors such as how well the question fits the user's interest, user's self-confidence on the quality of answers, and user's expected gains from answering the question.

\textit{User's expertise level on the question}. Even if a user has noticed a question and is willing to answer it, the user may not have the adequate knowledge to give a high-quality answer. That is the first and foremost reason that we need an expert recommendation approach in CQA.

Besides identifying the different aspects, we need to find a reasonable way to combine them to recommend real experts more comprehensively given a new question.
Unfortunately, the existing expert recommendation methods usually consider only the second, the last, or both the above aspects.
For example, most language models and topic models focus on recommending users who are most likely to answer the given question. However, a high possibility of a user answering a question does not necessarily mean the user would be able to provide a high-quality answer.
Many link analysis methods identify experts as the most authoritative users in a user-user network, where authoritativeness is a different concept from either relevance and user expertise. Therefore, the most authoritative users are not guaranteed to be willing to answer a question nor being able to give a high-quality answer.
Besides, some classification methods merely rely on the previous answer quality and some metadata features without considering users' topical distributions.
For this reason, a recommended expert by such methods may not want to answer a question even if the user is a good answerer in general.
Worse still, to the best of our knowledge, the first aspect has not been considered by any previous research efforts.
A promising research direction is to incorporate expert recommendation method with models that could effectively predict user behaviors, just like the prediction of check-ins in a Point of Interest recommendation problem~\cite{yin2015joint}.

\subsection{Coping with Dynamicity}

Currently, the vast majority of research efforts consider the expert recommendation problem in a static context, where they use a snapshot of users' previously asked or answered questions for the expert recommendation.
However, the real-world question answering websites are dynamic, with new users joining and leaving, users' interest changing, users' roles transforming, users' mutual interactions evolving, and the content on the website never stopping updating~\cite{yin2016spatio}.
Therefore, it is especially promising to develop methods that leverage the temporal information to make the expert recommendation methods adaptive in a dynamic context in a real-time fashion.

Currently, \textit{user availability} is the most commonly consider dynamic aspect for the expert recommendation problem.
Several studies used temporal features to estimate the availability of users for a given day~\cite{li2010routing,sung2013booming,chang2013routing} or for a specific time of the day~\cite{liu2011modeling,chang2013routing}.
For example, Sung et al.~\cite{sung2013booming} use all replies of users to train a sigmoid function and Chang et al.~\cite{chang2013routing} build binary classifiers using all responses of users within a fixed time frame (previous day).
Differs from the above work, Yeniterzi et al.~\cite{yeniterzi2015moving} use only the particular question-related replies to estimate availability.

Besides user availability, we identify two promising directions to cope with the dynamicity in CQA:

\textit{User interest drifts}.
Similar to general recommendation systems, the expert recommendation problem for CQA also faces the user interest drift issue~\cite{yin2016adapting}.
A straightforward solution is to include a decaying factor to suppress questions answered in remote history and focus on users' recent interest (reflected by their shifting answering behavior)~\cite{szpektor2013relevance}.
Although the topic drift issue has been studied in multiple areas such as social networks and mobile crowdsourcing~\cite{tong2014tcs}, it is almost an unexplored topic in the CQA context.
More factors such as fluctuations in user behaviors and more sophisticated time series prediction methods could be employed to gain better results.

\textit{Dynamic user expertise}.
Generally, users' expertise may not be consistent as well, as users' skills may improve over time and the quality of their answers may depend on various factors such as the users' status and other users' behaviors on a given question, not mentioning the impact from the evolution of the Q\&A community.
Pal et al.~\cite{pal2012evolution} analyze the evaluation of experts over time and show that estimating expertise using temporal data outperforms using static snapshots of the data.
In a very recent work~\cite{yeniterzi2015moving}, Yeniterzi et al. incorporate temporal information to model dynamic user expertise and apply two models, namely exponential and hyperbolic discounting models, to discount the effect of older records in calculating z-scores.
This method is still rather straightforward being equivalent to using a decaying factor. In particular, the z-scores are calculated for each time interval and then discounted according to its temporal distance from the question's hosting interval.
In the field of answer quality prediction, Szpektor~\cite{cai2013improving} use a more advanced set of temporal features calculated between the time at which a question and its replies are posted to improve prediction results. The similar method applies o the expert recommendation problem.

In summary, all the above dynamic aspects are suggesting an expert recommendation method suitable to self-evolve in an online fashion~\cite{tong2016online}.
However, none of the above methods is designed to be online-friendly, and it could take tremendous time to retrain the new model when new information becomes available, which is unacceptable in practice as most Q\&A systems in the real-world involves a massive amount of data overall.
Therefore, a promising research direction is to introduce novel methods that are capable of incrementally learning about users and continuously adapting their recommendation behaviors over time effectively and efficiently.
Besides the short of related work, we believe the dynamicity-related research for CQA is still at a preliminary stage, as most of the methods used are relatively simple and predict different aspects of consideration, such as user availability, user interest, and user expertise, separately. Moreover, they have not considered the possible correlations among these aspects.
Therefore, another potential point of research is to predict different aspects of features simultaneous using a single, comprehensive model for better results.
As an example, Tensor Factorization (TF)~\cite{yin2017sptf} may model the correlations among high-dimensional features better than Matrix Factorization (MF).

\subsection{Recommending Experts as a Collaborative Group}

Since a single user may not be able to well or fully address a given question, most methods would recommend either a ranked list or an unranked group of users instead of one user.
The recommended user group is expected to address the question better than a single user, and their answers are supposed to be better than most, or ideally all the other possible user groups that contain the same number of members when considered as a whole.
An ideal expert group should satisfy the following conditions.
First, the question must appeal to all group members so that they are likely to answer the question.
Second, the group members should be compatible with one another, so that the existence of one user in the group would not discourage another user to answer the question.
Third, it is desirable for the group members to complement one another in the knowledge domains required to address the question, given that users may have different sets of skills and differed level of expertise on different skill aspects~\cite{yin2011finding}.
Another benefit of group recommendation is the potential to make the recommending technique adaptive to specific scenarios and reduce the generation of redundant information in Q\&A communities.
For example, by capturing a global insight into an expert group, a method can automatically adjust the number of experts to recommend. In this way, the difficult questions may get more answers than easier ones.

To better answer a question, it is necessary to evaluate and select a competitive combination of potential answerers as a collaborative group.
Unfortunately, there is rarely any studies on this topic, and the group recommendation methods for traditional recommender system assume known user groups before making a recommendation~\cite{o2001polylens,ye2012exploring,gorla2013probabilistic}.
For example,
Pal et al. propose an expert group recommendation for the CQA~\cite{pal2015metrics}, aiming to find the experts from predefined communities to provide an answer.
Given a new question, the authors compute its similarity with the three aspects of features of each community, namely question features, user features, and community features.
Then, they use the two $k$-NN algorithms over the similarity metrics to build a vector of community scores for each community. 
Finally, they use linear regression, Borda Count, SVM Ranking, as well as the combinations of the above three methods to train a binary classifier for distinguishing the desired from the non-desired communities for the given question.
The issue with this method, as well as the traditional group recommendation methods, is that the users are evaluated separately and later put together to form a group.
Consequently, the workers' answers may not collectively better address the question from the requester's perspective as the second and third conditions above may not be well met.

Intuitively, the users who have frequently answers the similar question are likely to be compatible with one another~\cite{feng2018expert}.
A possible solution following this insight is to propose an expert recommendation scheme that aims at selecting the best subset (e.g. of a size of \textit{k}) of collaborative users by learning their co-occurrence in the same thread and topical expertise simultaneously.
The selection of a group of collaborative users could also borrow ideas from two closely related topics, namely optimal \textit{task decomposition}~\cite{tong2018slade} and \textit{user group evaluation}.
Task decomposition is the opposite approach of group formation, which aims to break the knowledge requirement into sub-requirements and find a user for every sub-requirement to compose the final group.
User group evaluation aims to set better heuristics to promote better recommendation results and answers to the given question.
On this aspect,  Chang et al.~\cite{chang2013routing} hypothesize that valuable question-answer threads are those where several people collaborate.
A natural application of this hypothesis in group expert recommendation is that, instead of aiming to maximize the requester's satisfaction on the given question, we could select the group of collaborative users in such a way that maximizes the long-term value to a broader audience of the thread.

\subsection{Leveraging External Information}

To address the cold start problem and the sparsity of data, especially, users with the low level of activity, it is crucial to leverage external data to facilitate the better expert recommendation in CQA.
The types of non-CQA data may vary depending on the external services and platforms.
Typically, they include the ``about me" description, homepage, blogs, micro-blogs, or social networking sites.
For example, many users make their GitHub homepages public in Stack Overflow, which could be an external source of information to estimate the users' expertise.

Until now, existing expert recommendation research for CQA only uses external information in a narrow scope in a relatively simplistic manner.
They mostly focus on users' social attributes including inter-relationship in social networks (either within or outside the Q\&A system)~\cite{atkinson2017redundancy}, and they either obtain user attributes through heuristic statistics outside of the model or combine users' social attributes with the original expert recommendation model by linear interpolation.
The limited related work includes: 
Srba et al.~\cite{srba2015utilizing} use non-CQA sources of data in topic-model-based approaches as a supplement for Q\&A activities in expertise estimation;
Zhao et al.~\cite{zhao2015cold} consider both topical interests and the ``following relations" between the users to build a user-to-user graph. The graph is then used in combination with past question-answering activities of users to evaluate and rank users.
Instead of using social attributes as heuristics to estimate user expertise, both Liu et al.~\cite{liu2014predicting} and Luo et al.~\cite{luo2014have} directly use users' social characteristics as additional features in addition to user expertise for the expert recommendation in CQA;
Zhao et al.~\cite{zhao2015expert} combine heuristics from two different social networks, i.e., social following and social friendship on Twitter and Quora for ranking answerers.

Besides the explicit links to external information sources for the users in a Q\&A system, we identify two promising fields of research that could avail the detection and use of external information by an expert recommendation method.
The first is account linkage techniques, which aim to identify and link to accounts of the same users on different systems such as websites. By automatically detecting the linked account of a user in CQA, the external information for this user could be efficiently extracted and utilized.
The second is cross-domain learning, represented by transfer learning techniques, which aims to utilize the information in the related or similar domains to help learn user models for a targeted domain.
Though great potentials in the Q\&A domain, both techniques have not yet currently introduced to the CQA.

\subsection{More Comprehensive Solutions}

Despite hybrid methods have considered multiple aspects of concern comprehensively, the research in this area is still at a preliminary stage as many of those methods simple combine the calculation results on different aspects as a weighted sum. Considering this deficiency, it is beneficial to develop more comprehensive methods.
To this end, we advocate several approaches beyond the existing techniques for the expert recommendation in CQA: factorization machines, ensemble learning, graph embedding, and deep learning models.
We will briefly discuss them in the following.

\textit{Factorization machines (FM)}~\cite{rendle2010factorization} is a matrix factorization based machine learning models similar to linear regression models.
It represents a generic approach that combines the generality of feature engineering with the superiority of factorization models in estimating interactions between the massive variables in a vast domain. FM model has the advantages of embedded variable interactions, reliable estimation of a linear number of parameters under high sparsity, and applicability to a variety of prediction tasks including regression, binary classification, and ranking.
All these advantages make FM a better replacement of traditional recommendation methods such as matrix factorization and tensor factorization for the expert recommendation in CQA.

\textit{Ensemble learning}~\cite{zhou2012ensemble} is a method of using multiple learning algorithms to obtain better performance than that obtainable by any of individual learning algorithm alone.
It generally includes parallel and the sequential ensemble learning and applies to various problems such as classification, regression, feature selection,
and anomaly detection.
Therefore, it could be potentially used to recommend experts in CQA.
Following the sequential or parallel ensemble paradigms, the candidate experts are either filtered by one learning algorithm after another or obtained by merging the recommended lists of experts by different algorithms.
Gradient tree boosting~\cite{friedman2001greedy}, represented by XGBoost~\cite{chen2016xgboost}, is another class of ensemble models that shines in many applications in recent years.
It is also promising yet unexplored in the context of CQA.

\textit{Graph embedding}~\cite{yan2007graph} is an effective and efficient way to solve the various graph analytics problems such as node classification, node recommendation, and link prediction while well overcoming the high computation and space cost issues of traditional graph analytics methods. It converts the graph data into a low dimensional space which maximally preserves the graph's properties and structural information.
In fact, graphs is an ideal form to represent the complicated interactions among various aspects of users, questions, and answers in CQA, especially when considering multiple aspects of concern and numerous information sources.
A benefit of applying the graph embedding approach is that various approximation algorithms and probabilistic solutions are readily available to address complex problems.

\textit{Deep learning}~\cite{yin2017spatial} has proven successful in many applications, and various deep learning models such as those based on autoencoders and neural autoregressive models have been applied to recommender systems~\cite{lin2017survey}.
Deep learning models have the advantage of utilizing multimodal heterogeneous features and thus has the potential of solving complex problems such as the expert recommendation problem on a large scale.
Convolutional neural network (CNN) is the only a deep learning model we are aware of that combines user feature representations with question feature representations to recommend experts for a given question in CQA~\cite{zheng2017deep}.

A closely related topic to the expert recommendation in CQA is question answering, which aims to find or generate answers to a given question automatically.
This topic is more classic and also heavily researched in the Q\&A research domain.
Other related research topics include question retrieval, answer quality/probability prediction, and expert finding in broader contexts.
In fact, though rare adoption for the expert recommendation problem, deep learning models have been widely applied for question answering in the domain of CQA.
Therefore, it could be a good idea to borrow and adapt the various sophisticated methods in these related domains to address the expert recommendation problem in CQA.

\section{Conclusions}
\label{conclusion}

In this survey, we focus on the expert recommendation problem, one of the most significant issues in Community question answering (CQA), and review the main techniques and state-of-the-art efforts on addressing the problem.
We have summarized and compared the existing methods in various aspects, including the datasets, input and output, evaluation metric, the covered aspects of concern, robustness over data distributions, and complexity, followed by discussing the advantages and shortcomings of these methods and pointing out the open issues and promising future research directions.
We hope this survey can help readers gain a quick and comprehensive understanding of the state of the art research in the expert recommendation in CQA and inspire more future research in this area.

%

\vspace{3mm}

\bibliographystyle{jcst}{
    \footnotesize
    \itemsep=-3pt plus.2pt minus.2pt
	\baselineskip=11pt plus.2pt minus.2pt

}


\vspace{5mm}

\noindent\parbox{8.3cm}{\parpic{\includegraphics[width=1in,height=1.25in]{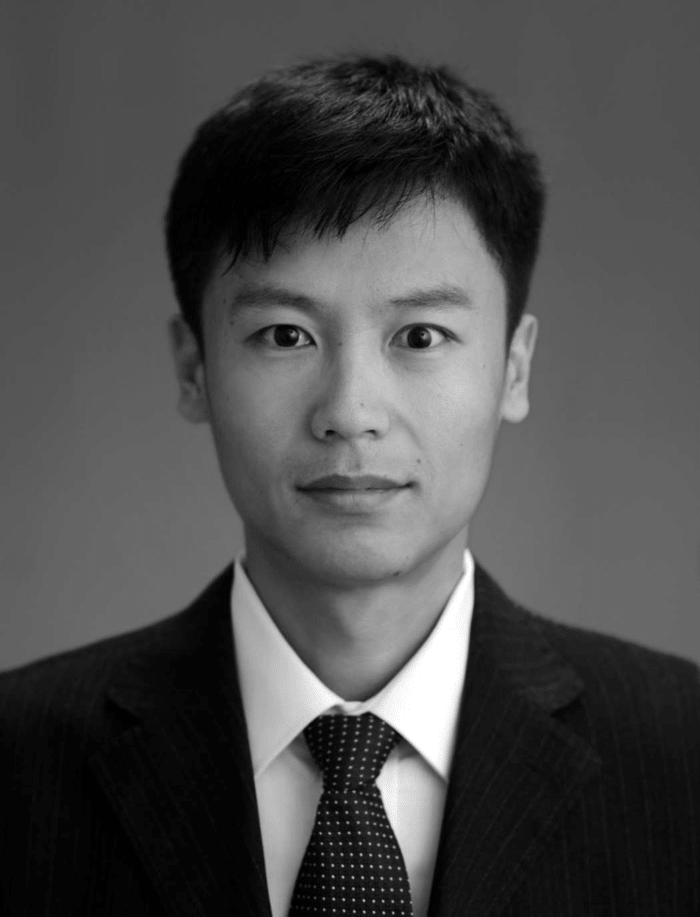}}{\small\quad {\bf Xianzhi Wang} is a research fellow with School of Information Systems, Singapore Management University, Singapore. He received his B.E. degree from Xi'an Jiaotong University, Xi'an, M.E. and Ph.D. degrees from Harbin Institute of Technology, Harbin, all in computer science in 2007, 2009, and 2014. His research interests include Internet of Things, data management, machine learning, and services computing. He received ARC Discovery Early Career Researcher Award (DECRA) in 2017 and IBM Ph.D. Fellowship Award in 2013.}\\[1mm]}

\noindent\parbox{8.3cm}{\parpic{\includegraphics[width=1in,height=1.25in]{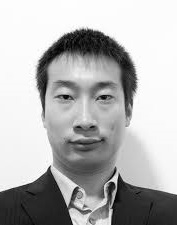}}{\small\quad {\bf Chaoran Huang} is currently a Ph.D. candidate at School of Computer Science and Engineering,
University of New South Wales, Sydney. He received
his B.E. degree from Tianjin
Polytechnic University, Tianjin, in 2014. His research interests include data mining, Internet of Things, and service-oriented computing.}\\[1mm]}

\noindent\parbox{8.3cm}{\parpic{\includegraphics[width=1in,height=1.25in]{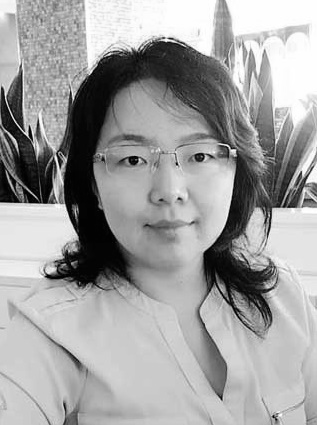}}{\small\quad {\bf Lina Yao} is a lecturer at School of Computer
Science and Engineering, University of New
South Wales, Sydney. She received her Ph.D. in computer
science from University of Adelaide, Adelaide, in 2014.
Her research interests include data mining and
machine learning applications with the focuses
on Internet of Things, recommender systems,
human activity recognition, and Brain Computer
Interface. She is the recipient of ARC Discovery
Early Career Researcher Award (DECRA) and
Inaugural Vice Chancellor's Women's Research
Excellence Award in 2015.}\\[1mm]}

\noindent\parbox{8.3cm}{\parpic{\includegraphics[width=1in,height=1.25in]{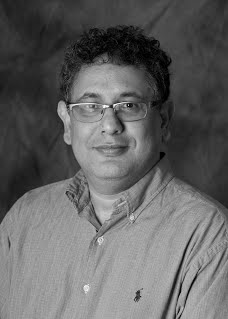}}{\small\quad {\bf Boualem Benatallah} is a scientia professor and research group leader at University of New
South Wales, Sydney. He received his Ph.D. in computer science from Grenoble University, Grenoble. His research interests include Web services composition, quality control in crowdsourcing services, crowdsourcing for vulnerability discovery, data curation, cognitive services engineering, and cloud services orchestration.}\\[1mm]}

\noindent\parbox{8.3cm}{\parpic{\includegraphics[width=1in,height=1.25in]{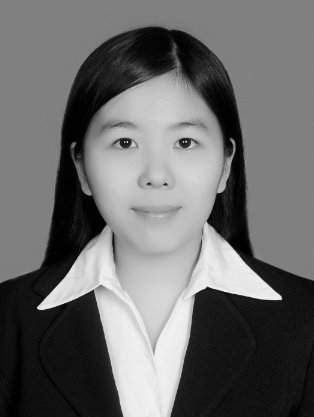}}{\small\quad {\bf Manqing Dong} is currently a Ph.D. candidate at School of Computer Science and Engineering, University of New South Wales, Sydney. She received her BE degree from Jilin University, Jilin, and her MSc in the City University of Hong Kong, Hongkong. Her research interests include anomaly detection, data mining, deep learning, statistical learning, and probabilistic graphical models. She is a student member of IEEE and ACM.}\\[1mm]}

\label{last-page}
\end{multicols}
\label{last-page}
\end{document}